\documentclass[aps,preprint,prb,showpacs]{revtex4-1}
\usepackage{epsf}
\usepackage{graphicx}
\usepackage{subfigure}

\usepackage{amsmath}

\begin{document}


\title{Surface and bulk TM polaritons in a linear magnetoelectric multiferroic with canted spins}


\author{V. Gunawan}
\email[]{slamev01@physics.uwa.edu.au}
\author{R.L. Stamps}
\affiliation{School of Physics M013, University of Western Australia, 35 Stirling Highway, Crawley, Western Australia 6009, Australia}


\date{\today}

\begin{abstract}
We present a theory for surface polaritons on ferroelectric-antiferromagnetic materials with canted spin structure. Canting is assumed to be due to a Dzyaloshinkii-Moriya interaction with the electric polarisation and weak ferromagnetism  directed in the plane parallel to the surface. Surface and bulk modes for a semi-infinite film are calculated for the case of transverse magnetic polarisation.  Example results are presented using parameters appropriate for BaMnF$_4$.  We find that the magnetoelectric interaction gives rise to "leaky" surface modes, i.e. pseudosurface waves that exist in the pass band, and that dissipate energy into the bulk of material. We show that these psuedosurface mode frequencies and properties can be modified by temperature, and application of external electric or magnetic fields.
\end{abstract}


\maketitle


\section{Introduction\label{intro}}
Magnetic polaritons are electromagnetic waves that travel in a  material with dispersion and properties modified through coupling to magnetic excitations\cite{camley82}. Polaritons can display a number of interesting and useful properties, including localization to surfaces and edges, and non-reciprocity\cite{harstein73,camley82,boardman,barnas86a}, whereby propagation frequency may not symmetric under direction reversal: i.e. $\omega(k)\neq\omega(-k)$. Surface polaritons at optical frequencies have received much attention in recent years, and appear in a number of different applications including detectors\cite{nylander}, biosensors\cite{liedberg} and microscopy\cite{keilmann98}.

Theoretical treatments for ferromagnetic polaritons \cite{kars78,harstein73} and simple antiferromagnets\cite{camley82,abraha96,camley98} were made several years ago. A most interesting class of polaritons are in multiferroic materials where magnetoelectric interactions couple magnetic and electric responses \cite{barnas86a,barnas86b,tarasenko00}.  A focus of theoretical work has been on bulk modes in linear magnetoelectric coupled media\cite{barnas86a,barnas86b}. Surface modes have also been discussed for the case of no applied external magnetic or electric fields, and neglecting canting of magnetic sublattices\cite{buchel86,tarasenko00}.

Modification and control of multiferroic surface polaritons  through external electric and magnetic fields is an intriguing prospect.  In the present paper we discuss in detail how temperature, electric and magnetic fields affect surface modes in canted spin multiferroics with linear magnetoelectric coupling. We allow for canting of magnetic sublattices, and show that canting is very important for understanding and manipulating surface mode frequencies. Most significantly, we show that the transverse magnetic field polarisation (TM) surface polariton excitations are in fact pseudo-surface modes characterized by complex propagation wavevectors whose imaginary parts are proportional to the strength of the magnetoelectric interaction.

Linear magnetoelectric coupling is believed to operate in multiferroics BaMnF$_4$\cite{tilley82} and FeTiO$_3$\cite{ederer08}. In order to make contact with previous work on bulk polaritons, we concentrate here on the BaMnF$_4$ system. The paper is organized as follows. The geometry and energy density of the system are considered in Section II where we also discuss the canting angle in relation to the magnetoelectric coupling.  In section III, susceptibilities are derived using Bloch  and Landau-Khalatnikov equations.  The electromagnetic problem is solved in Section IV and results given in Section V for surface and bulk modes on BaMn$_4$. In Section VI, effects on surface mode properties due to possible modifications of material parameters are discussed. Conclusions are given in Section VII.


\section{Geometry\label{geo}}
\begin{figure}[ht]
\includegraphics[width=7cm]{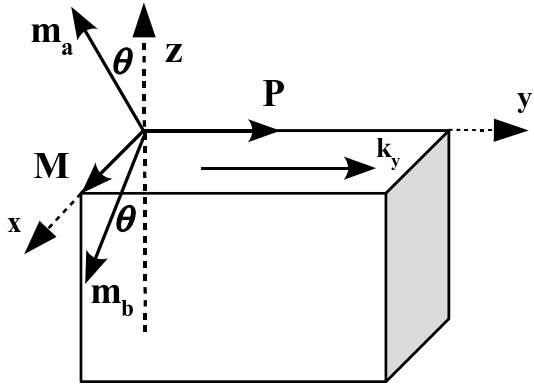}
\caption{\label{geometry}Geometry. Canting of two magnetic sublattices ($m_a$ and $m_b$) by an angle $\theta$ produces a weak ferromagnetism ($M$) along the $\hat{x}$ axis parallel to the surface. The spontaneous polarization ($\vec{P}$) is assumed to lie in a plane parallel to the surface. Propagation of the surface mode is along the $\hat{y}$ axis with wavenumber $\vec{k}_y$.}
\end{figure}

The geometry is sketched in Fig.\ref{geometry}. We consider a semi-infinite multiferroic film that fills the half space $z<0$.  The magnetic component of the multiferroic is a two sub-lattice antiferromagnet with uniaxial magnetic anisotropy. The two magnetic sub-lattices are allowed to cant in the $x-z$ plane  with  canting angle, $\theta$. We assume symmetric canting such that $\left|{\vec{{m_{a}}}}\right|=\left|{\vec{{m_{b}}}}\right|=M_{s}$.  The canting generates a weak ferromagnetism which is perpendicular to the spontaneous polarisation.  This configuration represents a Dzyaloshinkii-Moriya canting driven by spontaneous polarisation.  Both the weak ferromagnetic moment and spontaneous polarisation are constrained to lie in $x-y$ plane, parallel to the surface.  The magnetic easy axis is out-of-plane, along the $z$ direction. An external electric field is applied parallel to the spontaneous polarisation, and an external magnetic field is applied along the weak ferromagnet moment.

For the polariton propagation, we consider in this paper transverse magnetic (TM) polarization in which the magnetic part of the electromagnetic wave propagates parallel to the surface.  We consider only  surface modes traveling along the $\hat{y}$ direction, so that the magnetic component lies in $\hat{x}$ direction while the electric component has $E_y$ and $E_z$ components.

A fourth order Landau-Ginzburg energy density is assumed to describe the dielectric contribution to the energy:
\begin{equation}
\label{freelist}
F_{e}=\frac{1}{2}\zeta_{1}P^{2}_{y}+\frac{1}{4}\zeta _{2}P^{4}_{y}+\frac{1}{2}\Delta_{1}\left(P^{2}_{x}+P^{2}_{z}\right)+\frac{1}{4}\Delta_{2}\left(P^{4}_{x}+P^{4}_{z}\right)-P_{y}E_{y}.
\end{equation}
The first and second terms on the right hand of Eq.(\ref{freelist}) represent the  energy density for the $y$ component of the polarization with $\zeta_1$ and $\zeta_2$  dielectric stiffnesses.  The third and fourth terms represent the contribution of $x$ and $z$ polarization components with dielectric stiffnesses $\Delta_1$ and $\Delta_2$. The last term is the  external electric field applied parallel to the spontaneous polarization $P_y$.

The magnetic contribution to the  energy density is assumed to be of the form:
\begin{equation}
\label{freemag}
F_{M}=\lambda \vec{M}_{a}\cdot\vec{M}_{b}-\frac{K}{2}\left[(\vec{M}_{a}\cdot \hat{z})^{2}+(\vec{M}_{b}\cdot\hat{z})^{2}\right]-\left(M_{a}+M_{b}\right)_{x}H_{o}.
\end{equation}
The first term on the right of Eq.(\ref{freemag}) is an exchange energy with a strength $\lambda>0$.  The second term represents the anisotropy energy with anisotropy constant $K$ and the last term is the Zeeman energy from an external magnetic field.  Although the easy axis is out of plane, we assume that the symmetry of the canted antiferromagnet results in a negligible magnetization along the \textit{z }direction and therefore ignore demagnetization effects.

The weak ferromagnetic moment is denoted $M_x$, and the longitudinal component of the magnetization is $L_z$. These are defined by:
\begin{equation}
M_x=\left(\vec{m}_a+\vec{m}_b\right)_x=2M_s\sin\theta
\end{equation}
and
\begin{equation}
L_z=\left(\vec{m}_a-\vec{m}_b\right)_z=2M_s\cos\theta.
\end{equation}

A linear magneto-electric coupling is assumed of the form
{\setlength\arraycolsep{2pt}
\begin{eqnarray}
\label{ME}
&F_{ME}&=-\alpha P_{y}M_{x}L_{z}
\nonumber\\
& &=-\alpha P_{y}\hat{y}\cdot\left(\vec{m}_a\times\vec{m}_b\right)-\alpha P_{y}\left[\left(m_a\right)_x\left(m_a\right)_z-\left(m_b\right)_x\left(m_b\right)_z\right]
\end{eqnarray}}
where $\alpha$ is the magneto-electric coupling constant.  This energy governs the canting of the magnetic sub-lattices.

The canting angle  $\theta$ is determined by minimizing the magnetic and magnetoelectric  energies with respect to  $\theta $. Minimizing, we arrive at the condition
\begin{equation}
\label{sdteq2}
H_{o}\cos \theta -\frac{1}{2}\mathit{KM}_{s}\sin 2\theta +2\alpha P_{y}M_{s}\cos 2\theta +\lambda M_{s}\sin 2\theta =0.	
\end{equation}
In the absence of an external magnetic field, Eq.(\ref{sdteq2}) simplifies to
\begin{equation}
\label{sdt}
\tan (2\theta )=\frac{4\alpha P_{y}}{K-2\lambda}.
\end{equation}
Note that a positive magnetoelectric constant describes
a weak ferromagnetism $M_x$ aligned along $-x$.  

The canting angle depends on the equilibrium magnitude of $P_y$, and this is found by minimizing the dielectric and magnetoelectric energies. Requiring  $\frac{\partial }{\partial P_y}(F_E+F_{ME})=0$ results in
\begin{equation}
\label{eqel}
\zeta _{1}P_{y}+\zeta _{2}P^{3}_{y}-2\alpha M_{s}^{2}\sin{2\theta} -E_{y}=0.	
\end{equation}

Lastly, the magnitude of $M_S$ depends on temperature. In mean field, the magnitude can be written in terms of the Brillouin function $B\left(\eta\right)$ as
\begin{equation}
\label{bril}
M_{s}=M_{s}(0)B_{s}(\eta)
\end{equation}
where
\begin{equation}
\eta =\frac{g\mu _{B}S}{k_{B}T}\left[-\lambda M_{s}\cos 2\theta
+\mathit{KM}_{s}\cos ^{2}\theta +2\alpha P_{y}\sin 2\theta +H_{o}\sin
\theta \right].	
\end{equation}
The spontaneous polarization, magnetization and canting angle are calculated by solving simultaneously Eqs.(\ref{sdteq2}), (\ref{eqel}) and (\ref{bril}).  Solution for general angles is done numerically using root finding techniques for coupled transendental equations.


\section{Dynamic Susceptibility\label{dynamic}}

In order to solve the electromagnetic boundary value problem for the surface and bulk polariton modes, we need constituitive relations for the dielectric and magnetic responses. We consider linear response and calculate the permitivity and permeability using equations of motion derived from Eqs.(\ref{freelist}),(\ref{freemag})and (\ref{ME}). The equations of motion for the magnetic response are given by Bloch equations,
\begin{equation}
\dot{\vec{M}}=\gamma \vec{M}\times \left(\frac{-\partial }{\partial\vec{M}}(F_M+F_{ME})\right)
\end{equation}
where  $\gamma $ is the gyromagnetic ratio. The equations of motion for the polarization response are given by Landau-Khalatnikov equations
\begin{equation}
\ddot{\vec{P}}=-f\frac{\partial}{\partial \vec{P}}(F_E+F_{ME})
\end{equation}
where $f$ is the inverse of phonon mass.  
 
The set of dynamic equations appropriate for the polarizations of TM modes are,
\begin{equation}
\label{magdyn1}
-i\omega m_x = \left(\omega_a\cos\theta+2\omega_{me}\sin\theta\right)l_y
\end{equation}
\begin{eqnarray}
&-i\omega l_y =& \left(2\omega_{ex}\cos\theta-\omega_a\cos\theta-4\omega_{me}\sin\theta\right)m_x
\nonumber\\
& &+\left(2\omega_{ex}\sin\theta+4\omega_{me}\cos\theta-\omega_a\sin\theta+\omega_o\right)l_z
\nonumber\\
& &+4\gamma\alpha M^{2}_{s}\cos{2\theta}p_y+2\gamma M_sh_x\cos\theta
\end{eqnarray}
\begin{equation}
\label{magdyn2}
	-i\omega l_z = -\left(2\omega_{ex}\sin\theta+2\omega_{me}\cos\theta+\omega_o\right)l_y
\end{equation}
\begin{equation}
\label{eldyn}
	\frac{\omega^2}{f}p_y = \left(\zeta_1+3P^{2}_{o}\zeta_2\right)p_y-2\alpha M_s\left(m_x\cos\theta+l_z\sin\theta\right)-e_y
\end{equation}
and
\begin{equation}
	\frac{\omega^2}{f}p_z = \left(\Delta_1+\Delta_{2}P^{2}_{o}\right)p_z-e_z.
\end{equation}

The notation used above is in units of frequency and defined as: $\omega_a=\gamma K M_s$ is the magnetic anisotropy, $\omega_{ex}=\gamma\lambda M_s$ is the exchange, $\omega_{me}=\gamma\alpha P_oM_s$ is the magnetoelectric coupling and $\omega_o=\gamma H_o$ is the external magnetic field.

Equations (\ref{magdyn1}) to (\ref{magdyn2}) are coupled to (\ref{eldyn}) through ME susceptibilities $\chi^{me}$ and $\chi^{em}$ defined as
\begin{equation}
	\vec{m}=\chi^m\vec{h}+\chi^{me}\vec{e} \qquad \textrm{and}  \qquad \vec{p}=\chi^e\vec{e}+\chi^{em}\vec{h}.
\end{equation}

The relevant magnetic susceptibilities for TM modes are given by
\begin{equation}
\label{sucept1}
	\chi _{x}^{m}=\frac{1}{2\pi}\omega_{s}\left(\omega _{a}\cos ^{2}\theta +\omega_{me}\sin 2\theta
\right)\left\{\frac{C_{\mathit{mx}}}{\left(\acute{\omega}_{m}^{2}-\omega^{2}\right)}+\frac{C_{\mathit{ex}}}{\left(\acute{\omega}_{\mathit{ey}}^{2}-\omega ^{2}\right)}\right\}
\end{equation}
where $\omega_s=\gamma 4\pi M_s$.
The electric susceptibilities are
\begin{equation}
	\chi _{y}^{e}=f\left\{\frac{C_{\mathit{ey}}}{\left(\acute{\omega}_{\mathit{ey}}^{2}-\omega
^{2}\right)}+\frac{C_{\mathit{my}}}{\left(\acute{\omega}_{m}^{2}-\omega ^{2}\right)}\right\},
\end{equation}
 \begin{equation}
\chi _{z}^{e}=\frac{f}{\left(\omega _{\mathit{ez}}^{2}-\omega^{2}\right)}	.
\end{equation}
The magnetoelectric susceptibility is
\begin{equation}
\label{sucept2}
	\chi _{\mathit{xy}}^{\mathit{me}}=\chi_{\mathit{yx}}^{\mathit{em}}=\frac{-C_{\mathit{me}}}{\left(\acute{\omega}^{2}_{ey}-\omega^{2}\right)}+\frac{C_{\mathit{me}}}{\left(\acute{\omega}^{2}_{m}-\omega ^{2}\right)}.
\end{equation}

The frequencies  $\acute{\omega }_{\mathit{ey}}$ and  $\acute{\omega}_{m}$ are defined as $\acute{\omega}_{\mathit{ey}}^2=\omega_{\mathit{ey}}^2+\delta $ and  $\acute{\omega }_{m}^2=\omega _{m}^2-\delta$, where  $\delta$ is expressed in the form
\begin{equation}
	\delta =\frac{1}{2}\left\{\left[\left(\omega _{\mathit{ey}}^{2}-\omega_{m}^{2}\right)^{2}-4\Omega _{C}^{4}\right]^{1/2}-\left(\omega_{\mathit{ey}}^{2}-\omega_{m}^{2}\right)\right\}
\end{equation}
where
\begin{equation}
\label{omegC}
\Omega _{c}^{4} = C_c\omega_s\cos 2\theta
\left(2\omega _{\mathit{ex}}\sin ^{2}\theta -\omega _{a}\cos ^{2}\theta+\omega _{o}\sin \theta \right)	
\end{equation}
with  $C_c=\frac{2}{\pi}\alpha ^{2}M_{S}^{2}f$.

$\omega _{\mathit{ey}}$ is the frequency of the soft phonon along the spontaneous polarization and  $\omega _{m}$ is the magnetic resonance frequency:
\begin{equation}
	\omega _{m}^{2}={\tilde {{\omega }}_{\mathit{afm}}}^{2}+\Omega_{\mathit{me}}^{2}+\Omega _{o}^{2}.
\end{equation}

Here  $\tilde {{\omega }}_{\mathit{afm}}$ is the antiferromagnet resonance, 
\begin{equation}
	{\tilde {{\omega }}_{\mathit{afm}}}^{2}=\omega_{a}\left(\omega _{a}-2\omega _{\mathit{ex}}\right)\cos ^{2}\theta+2\omega _{\mathit{ex}}\left(2\omega _{\mathit{ex}}-\omega_{a}\right)\sin ^{2}\theta
\end{equation}

 $\Omega_{\mathit{me}}$ is related to the magneto-electric interaction,
\begin{equation}
\Omega _{\mathit{me}}^{2}=8\omega _{\mathit{me}}^{2}+2\omega_{\mathit{me}}\left(\omega _{a}+2\omega _{\mathit{ex}}\right)\sin2\theta
\end{equation}

and $\Omega_{o}$ is related to the external magnetic field,
\begin{equation}
	\Omega _{o}^{2}=\omega _{o}\left[\omega _{o}+6\omega_{\mathit{me}}\cos \theta +\left(4\omega _{\mathit{ex}}-\omega_{a}\right)\sin \theta \right].
\end{equation}

The frequencies $\omega _{a}=\gamma \mathit{KM}_{s}$, $\omega_{\mathit{ex}}=\gamma \lambda M_{s}$, $\omega _{\mathit{me}}=\gamma\alpha P_{o}M_{s}$ and $\omega _{o}=\gamma H_{o}$ represent
contributions from energies associated with the magnetic anisotropy, exchange, magneto-electric coupling and external field respectively.  The frequency $\omega_{ez}$ is the phonon frequency along the $\hat{z}$ direction.

Other parameters in Eq. (\ref{sucept1}) to (\ref{sucept2}) are defined as
\begin{equation}
	C_{\mathit{mx}}=\left(\omega _{\mathit{ey}}^{2}-\acute{\omega}_{m}^{2}\right)/{\left(\acute{\omega }_{\mathit{ey}}^{2}-\acute{\omega}_{m}^{2}\right)}
\end{equation}
\begin{equation}
	C_{\mathit{ex}}=\delta /{\left(\acute{\omega
}_{\mathit{ey}}^{2}-\acute{\omega
}_{m}^{2}\right)}
\end{equation}
\begin{equation}
	C_{\mathit{ey}}=\left(\acute{\omega }_{\mathit{ey}}^{2}-\omega
_{m}^{2}\right)/{\left(\acute{\omega }_{\mathit{ey}}^{2}-\acute{\omega}_{m}^{2}\right)}
\end{equation}
\begin{equation}
	C_{\mathit{my}}=\delta /{\left(\acute{\omega}_{\mathit{ey}}^{2}-\acute{\omega}_{m}^{2}\right)}
\end{equation}
and
\begin{equation}
\label{omegAF}
	C_{\mathit{me}}=C_{\alpha f}\omega_{s}\cos 2\theta \left(\omega_{a}\cos \theta +2\omega _{\mathit{me}}\sin \theta\right)/{\left(\acute{\omega}_{\mathit{ey}}^{2}-\acute{\omega}_{m}^{2}\right)},
\end{equation}
where $C_{\alpha f}=\frac{1}{\pi}\alpha M_sf$.

From the expression of the susceptibilities above, it can be seen that the applied magnetic field  directly influences the  susceptibilities through  $\Omega _{o}$. By way of contrast, the applied electric field
 changes the susceptibilities indirectly by affecting the magnitude of the spontaneous polarisation.


\section{Theory for bulk bands and surface modes}
Dispersion relations for the bulk modes are obtained by solving the electromagnetic Maxwell equations
\begin{align}
\label{max}
	\nabla\times\vec{E}=-\frac{1}{c}\frac{\partial\vec{B}}{\partial{t}}\\
	\nabla\times\vec{H}=\frac{1}{c}\frac{\partial\vec{D}}{\partial{t}}\nonumber
\end{align}
where the fields $\vec{D}$ with $\vec{E}$ and $\vec{B}$ with $\vec{H}$ in TM modes are connected through constitutive equations
\begin{equation}
	B_{x}=\mu_{x}H_{x}+4\pi\chi^{me}E_{y},\qquad D_{y}=\epsilon_{y}E_{y}+4\pi\chi^{me}H_{x} \qquad \textrm{and}  \qquad D_{z}=\epsilon_{z}E_{z}
\end{equation}
with permeability and dielectric functions defined as $\mu_{x}=1+4\pi\chi_{x}^{m}$ and $\epsilon_{y}=1+4\pi\chi_{y}^{e}$.

Plane waves are assumed for bulk traveling modes of the form $\vec{E},\vec{H}\sim e^{i\left(k_{y}y+k_{z}z-\omega t\right)}$.
Substitution into the Maxwell equations provides an equation for the bulk mode dispersion:
\begin{equation}
\label{bulk}
	\epsilon _{y}(\epsilon_{z}\mu_{x}\left(\frac{\omega}{c}\right)^{2}-k_{y}^{2})-\epsilon_{z}\left(4\pi\chi^{me}\frac{\omega}{c}\right)^{2}=0.
\end{equation}
The dispersion relation has solutions determined by the zeroes of the dielectric constant $\epsilon_{z}$, zeroes of the function $f(\mu,\epsilon)=\mu _{x}\epsilon_{y}-\left(4\pi\chi ^{me}\right)^{2}$. Equation (\ref{bulk}) diverges at the pole of dielectric constant $\epsilon _{z}$,  $\epsilon _{y}$ and $\chi^{me}$.

The dispersion relation for surface modes is calculated by assuming surface localized plane wave solutions of the form:
\begin{equation}
\label{mat}
	\vec{E},\vec{H}\sim e^{\beta z}e^{i\left(k_{y}y-\omega t\right)}\qquad
\textrm{for $z<0$}
\end{equation}
and
\begin{equation}
\label{vac}
	\vec{E},\vec{H}\sim e^{-\beta _{o}z}e^{i\left(k_{y}y-\omega t\right)}\qquad
\textrm{for $z>0$}.
\end{equation}

Substitution of Eqs. (\ref{mat}) and (\ref{vac}) into Eqs. (\ref{max}) provide an implicit relation for the attenuation constant in the material, 
\begin{equation}
\label{betamed}
	\epsilon _{z}\left(\beta +i4\pi\chi ^{\mathit{me}}\frac{\omega}{c}\right)^{2}=\epsilon _{y}k_{y}^{2}-\epsilon _{y}\epsilon _{z}\mu_{x}\left(\frac{\omega }{c}\right)^{2},
\end{equation}
and an explicit relation for the attenuation constant in the vacuum:
\begin{equation}
\label{betavac}
	\beta _{o}^{2}=k_{y}^{2}-\left(\frac{\omega}{c}\right)^{2}.
\end{equation}.


\begin{figure}[ht]
\begin{center}
\subfigure[\label{reldisp}Dispersion relation.]{
\includegraphics[width=5.5cm,angle=90]{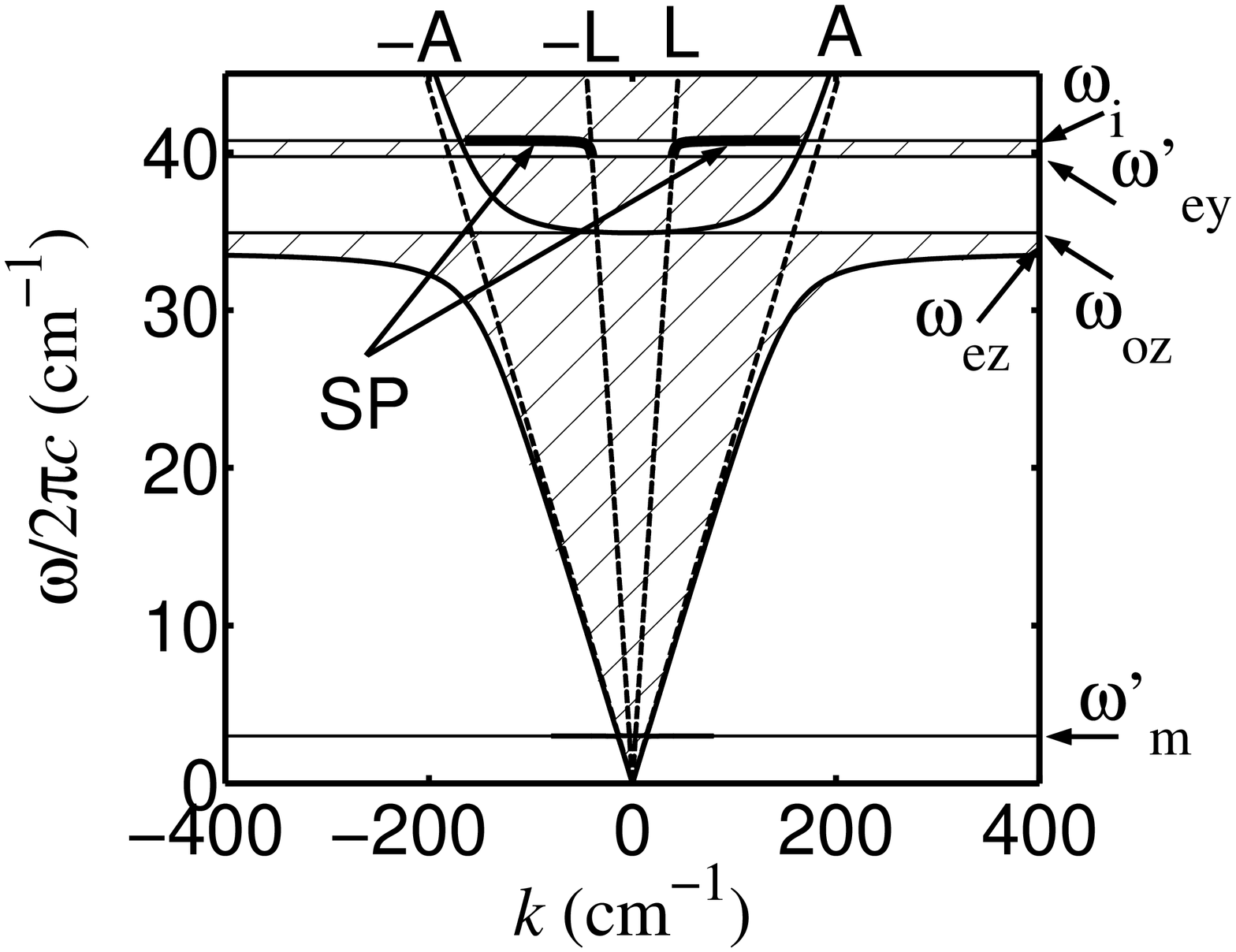}}
\subfigure[\label{window}'Window' where surface modes exist.]{
\includegraphics[width=6.5cm,height=5.5cm]{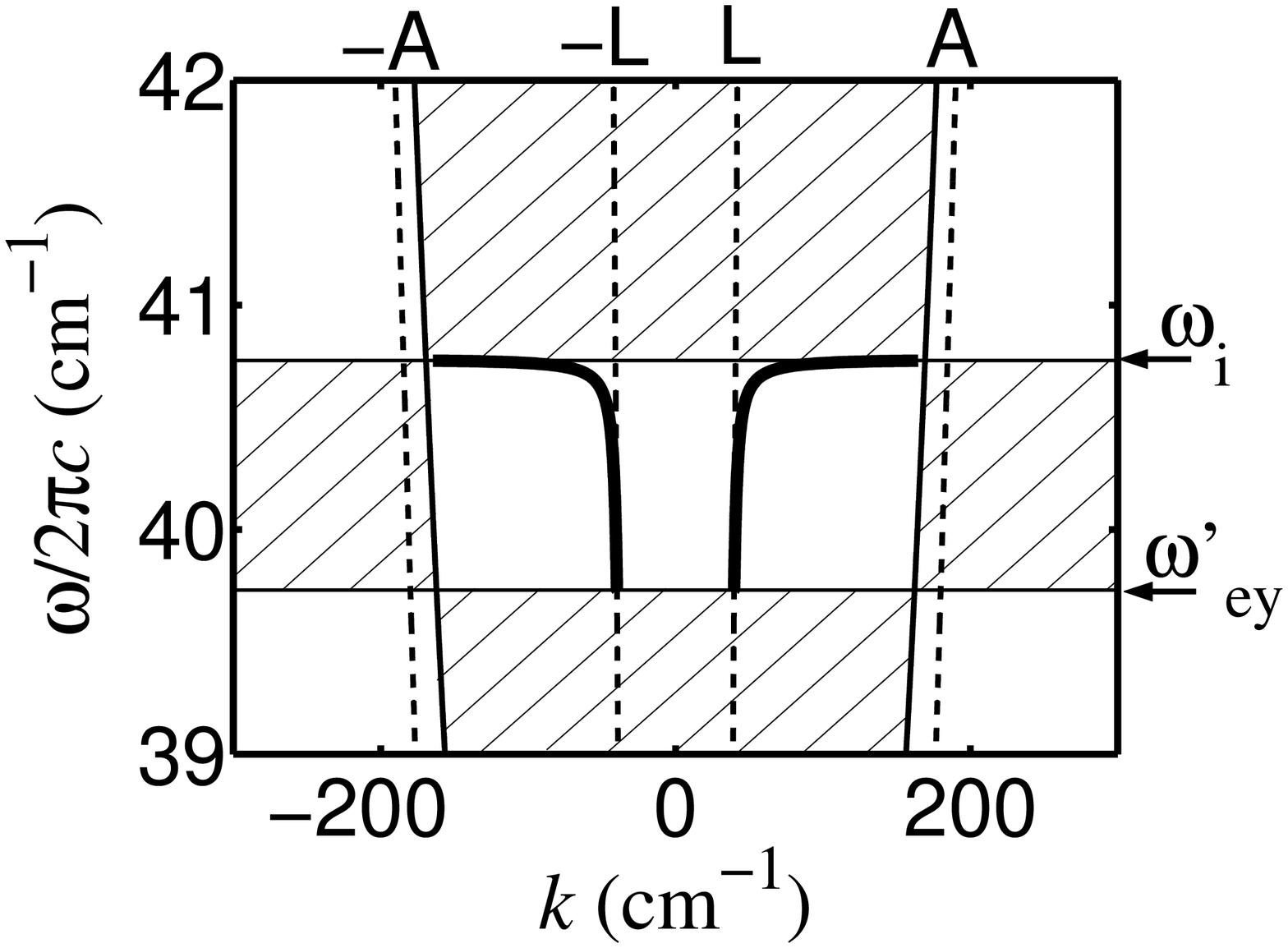}}
\subfigure[\label{slit}Narrow gap at $\omega_i$.]{
\includegraphics[width=7cm]{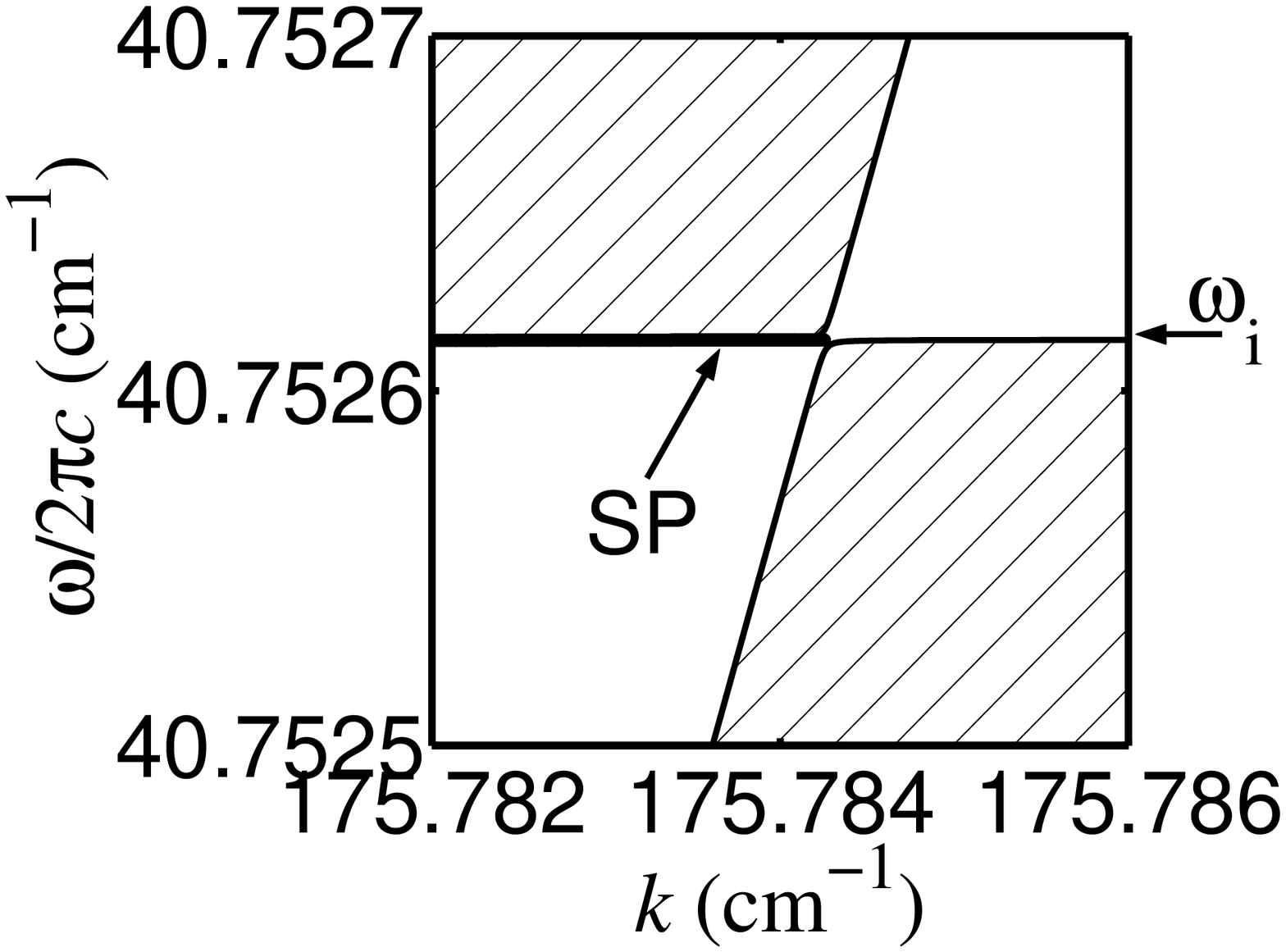}}
\subfigure[\label{slit40}Narrow gap with bigger coupling.]{
\includegraphics[width=6.5cm,height=5.1cm]{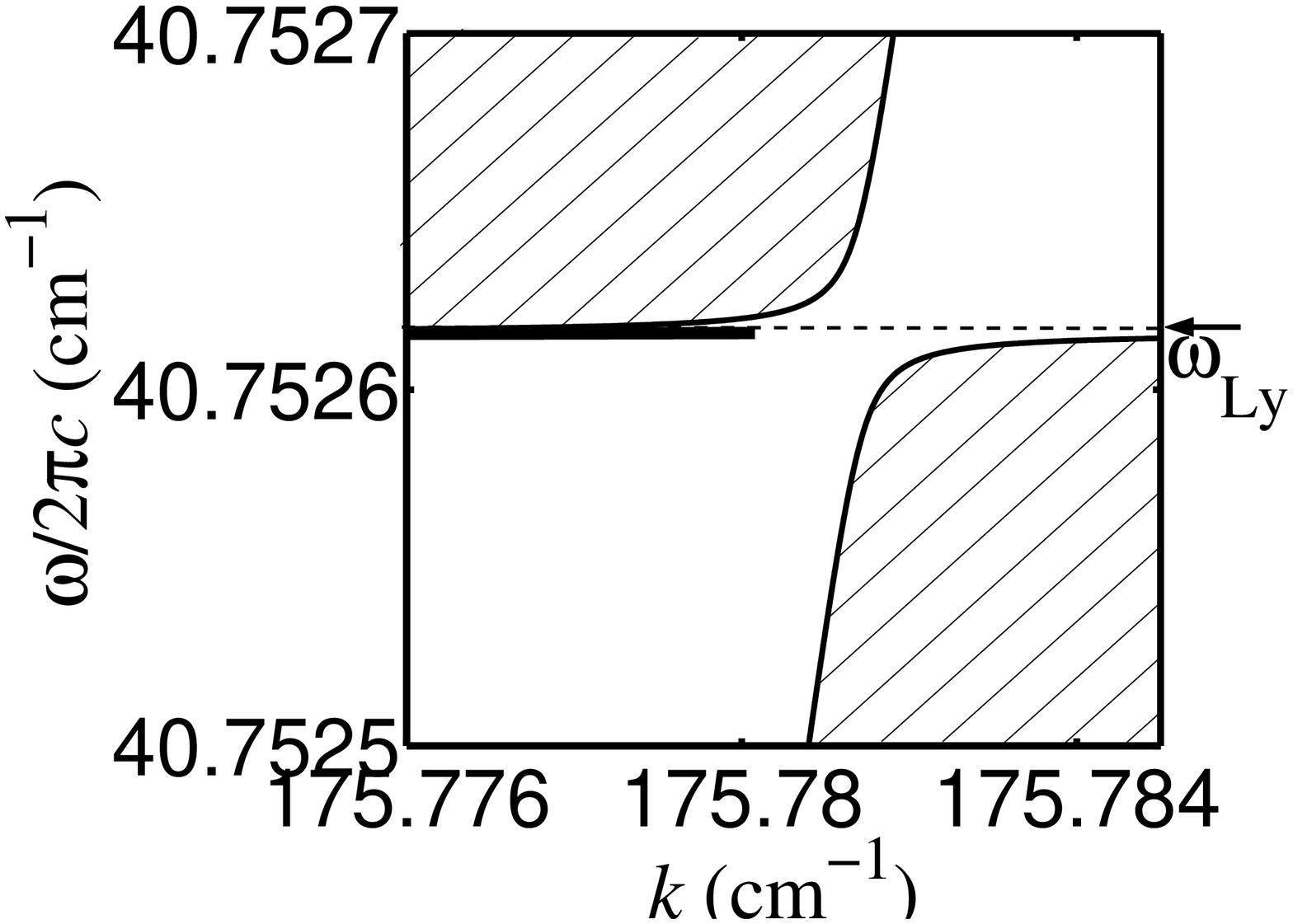}}
\caption{Dispersion relation without external field.  In (a) he dispersion relation without the external field is shown. The surface modes are indicated by{\textquotedblleft}SP{\textquotedblright}. The shaded regions represent bulk bands, which are limited by frequencies  $\acute{{\omega
_{m}}}$,  $\omega _{\mathit{ez}}$,  $\omega _{\mathit{oz}}$, $\acute{{\omega _{\mathit{ey}}}}$ and  $\omega _{i}$. In (b) The ``window'' where the surface modes exist is shown. In (c) a narrow gap around $\omega_i$ is expanded.  In (d) the narrow gap is wider when the ME coupling is increased by a factor of 10.}
\end{center}
\end{figure}

\begin{figure}[ht]
\begin{center}
\subfigure[\label{realbeta}Re($\beta$).]{
\includegraphics[width=7.1cm]{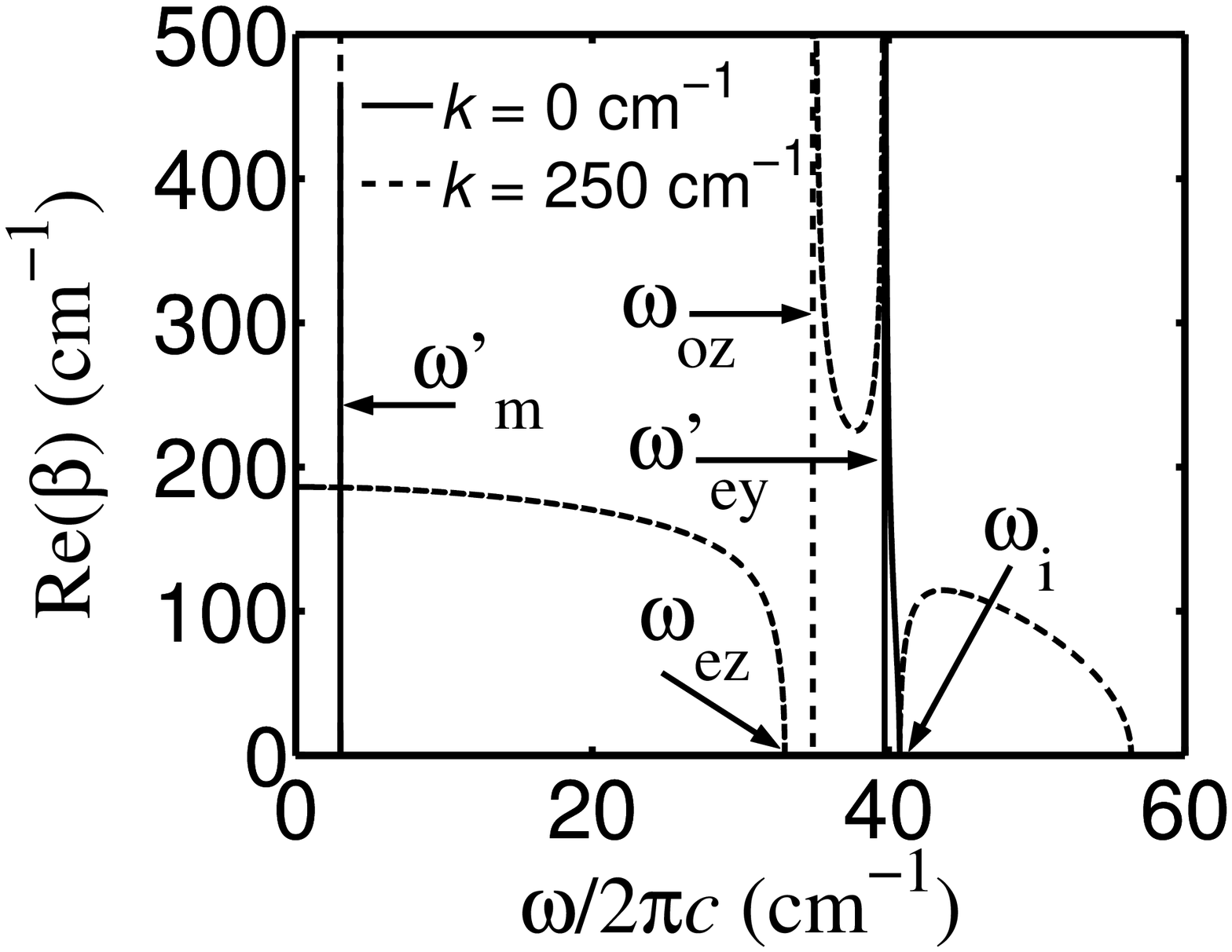}}
\subfigure[\label{imagbeta}Im($\beta$).]{
\includegraphics[width=7cm]{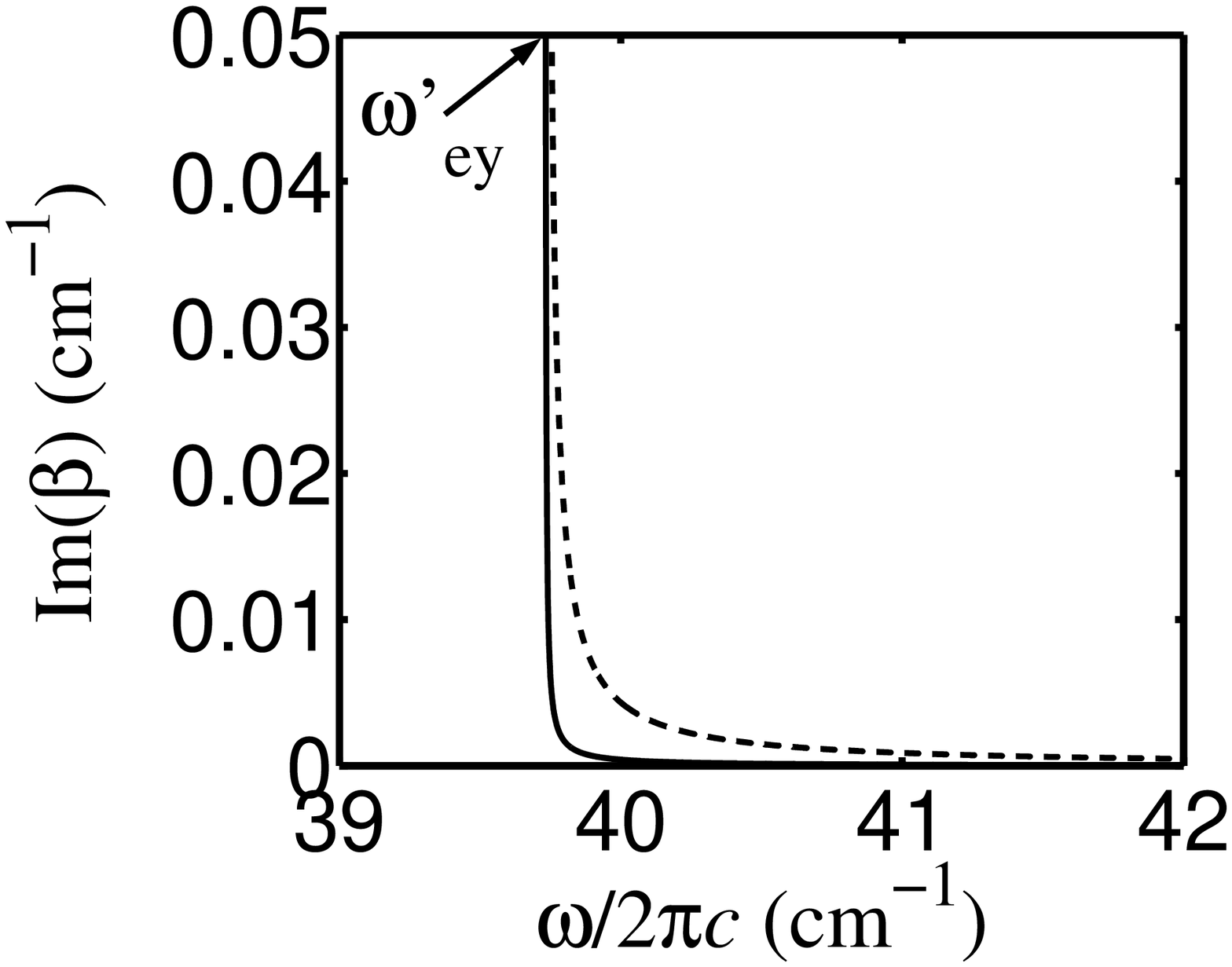}}
\caption{Attenuation constant as a function of frequency.  In (a) the real part of attenuation constant is shown for two values of wavevector in the absence of any external fields.  The solid line represents  k = 0  cm$^{-1}$, while the dashed line corresponds to  k = 250 cm$^{-1}$ .  In (b) the imaginary part of attenuation constant is shown for two values of ME coupling. The solid line is for $\alpha=1.42\times10^{-5}$ cm$^2$/statC, and the dashed line represents the coupling $\alpha=1.42\times10^{-4}$ cm$^2$/statC.}
\end{center}
\end{figure}
An implicit solution for the surface wave frequencies is found by matching the solutions in Eqs.(\ref{mat}) and (\ref{vac}) at \textit{z}=0 using electromagnetic boundary conditions. The unique conditions are continuity of tangential $\vec{H}$, $\vec{E}$ and continuity of normal $\vec{D}$. When satisfied, the following dispersion relation results:
\begin{equation}
\label{surf}
	\left(\beta +i4\pi\chi ^{\mathit{me}}\frac{\omega }{c}\right)+\epsilon_{y}\beta _{o}=0.
\end{equation}
Note that the terms in Eq.(\ref{surf}) do not have an odd multiple of wave vector $k$, and so the solutions should be reciprocal in the sense that $\omega\left(k\right)=\omega\left(-k\right)$.  We also see that the existence of surface modes strongly depends on the value of $\epsilon _{y}$, since the solution of Eq.(\ref{surf}) for surface
modes can only be found when the value of  $\epsilon _{y}$ is negative.


\section{Results from numerical calculations}
We now illustrate the preceding theory for the material BaMnF$_4$. Parameters appropriate for BaMnF$_4$ were determined as follows.  Measured values of the weak ferromagnetism\cite{scott79}, $M_{c}=18.347$ Oe, and a canting angle\cite{venturini} of 3 mrad inserted into the relation $M_{c}=2M_{s}\sin\theta $, yield the magnetisation of the sub-lattices $M_{s}=3.054\times10^{3}$ Oe.  Following the approximations by Holmes\cite{holmes69} for the exchange field, $H_{E}$= 50T, and the relation $H_{E}=\lambda M_{s}$, the exchange constant is  $\lambda $=163.72.  Using the measured value of the magnetic resonance frequency\cite{samara76}, $\omega_r$ = 3 cm$^{-1}$, in the relation $\omega_r=\gamma^{2}K(K-2\lambda)M_s^2$, we obtain the anisotropy constant \textit{K}=0.337.

The linear magnetoelectric coupling $\alpha$ is obtained by using the calculated spontaneous polarisation\cite{keve71} $P_{o}=3.45\times10^{4}$ statC/cm\textsuperscript{2} in  Eq.(\ref{sdt}), yielding $\alpha = 1.42\times 10^{-5}$ cm$^2$/statC.  The Ginzburg-Landau constant $\zeta_{1}$ and  $\zeta _{2}$ are approximated by solving simultaneously Eq.(\ref{eqel}) and  $\left(\zeta _{1}+3\zeta_{2}P_{o}^{2}\right)f=\omega ^{2}_{\mathit{ey}}$, the soft phonon frequency along the spontaneous polarisation. With the inverse phonon mass\cite{barnas86a} as $f = 2.7\times10^{24}$ statC\textsuperscript{2}/g cm\textsuperscript{3} and transversal phonon frequency\cite{samara76} $\omega _{\mathit{ey}}$= 7.73 THz, this gives $\zeta _{1}$=-10.528 g cm\textsuperscript{3}/statC\textsuperscript{2}s\textsuperscript{2} \ and  $\zeta _{2}=1.934\times10\textsuperscript{-8}$ erg cm$\textsuperscript{5}$/statC$\textsuperscript{4}$.  The suceptibility $\chi^e_z$ is calculated by using the transverse phonon frequency for $z$ polarization\cite{barnas86a}, $\omega_{ez}= 33.7$ cm$^{-1}$.  

The parameters $C_c$ and $C_{\alpha f}$ in Eq.(\ref{omegC}) and (\ref{omegAF}) are converted to frequency units in cm$^{-1}$. These become:
\begin{align}
\nonumber	C_c=\frac{2}{\pi}\frac{\alpha^2\mu_{o}M^2_sf}{(2\pi c)^2}\\
	C_{\alpha f}=\frac{1}{\pi}\frac{\alpha\mu_oM_{s}f}{(2\pi c)^2}
\end{align}
with $M_s$ in A/m, $\alpha$ in m$^2$/C, $f$ in C$^2$/kg m$^3$ and \textit{c} in cm/s.

Solutions of Eq.(\ref{bulk}) and (\ref{surf}) are plotted in Fig.(\ref{reldisp}) for the case with  no applied fields present.  There are two gaps in the bulk region which are created by the poles and zeros of Eq.(\ref{bulk}).  First, as illustrated in Fig.(\ref{slit}), a very narrow gap is located at the  frequency $\omega_i$ around 41 cm$^{-1}$, created by the magneto-electric interaction\cite{barnas86a}.  This gap is associated with zeros in  $f(\mu,\epsilon)=\mu _{x}\epsilon _{y}-\left(4\pi\chi ^{\mathit{me}}\right)^{2}$. The pole in the bulk modes is due to a zero of the dielectric constant  $\epsilon _{y}$.  This gap is strongly dependent on the ME susceptibility, $\chi^{me}$, and disappears when $\chi^{me}$=0.  The width of the gap is approximately proportional to $(\chi^{me})^2$.  Thus the gap becomes wider with larger ME coupling. This increase is illustrated in Fig.(\ref{slit40}) where the ME coupling has been increased by a factor of ten.  

A second gap exists near the magnetic frequency $\acute{\omega }_{m}$${\cong}$3 cm\textsuperscript{{}-1 }, and occurs at zero of $f(\mu)=\mu_{x}\epsilon_{y}-\left(4\pi\chi^{\mathit{me}}\right)^{2}$ at the magnetic frequency  $\acute{\omega }_{m}$.  The other boundaries for the bulk regions are determined by the attenuation constant.

Using the attenuation constants, we obtain a narrow window between transversal and longitudinal phonon frequencies, $\omega'_{ey}$ and $\omega_{Ly}$, associated with the pole and zero value of $\epsilon_y$ (see Fig.\ref{window}).  In this figure, since the ME coupling is weak, hence the frequency $\omega_{Ly}$ is very slightly below the induced frequency $\omega_{i}$.  Since the value of $\epsilon_y$ between these two frequencies is negative, surface modes can be obtained inside this narrow window.  The surface modes start from the crossing between the lightline  $\left(\omega =ck\right)$ and the resonance frequency  $\acute{\omega }_{\mathit{ey}}$ and terminate at the longitudinal phonon frequency $\omega_{Ly}$.  The frequency $\omega_{Ly}$ can be approximated as
\begin{equation}
\label{WL}
\omega_{Ly}=\frac{1}{2}\left\{\left(\omega^2_{ey}+\omega^2_m+f\right)+\left[\left(\omega^2_{ey}+\omega^2_m+f\right)^2-4\left(\acute{\omega}^{2}_{ey}\acute{\omega}^{2}_m+\omega^2_m f\right)\right]^{1/2}\right\}
\end{equation}
and is indicated in Fig.\ref{slit40}.  Since the surface modes terminate at the longitudinal phonon frequency, the gap around $\omega_i$ does not influence the surface modes.

Interestingly, from the expression for the attenuation constant of Eq.(\ref{betamed}), and the dispersion relation of surface modes in Eq.(\ref{surf}), it was only possible to satisfy the boundary conditions with a complex \textit{${\beta}$}. The resulting mode is not a true surface mode but is instead a pseudo-surface wave\cite{lim}. Because the solution for the attenuation constant in Eq.(\ref{betamed}) is complex, the attenuation constant for the material sample consists of real and imaginary part as illustrated in Fig.\ref{realbeta} and \ref{imagbeta}.  

The imaginary part of \textit{${\beta}$} is
\begin{equation}
	\beta _{i}=-i4\pi\chi ^{\mathit{me}}\frac{\omega}{c}.
\end{equation}
The positive value of the real part defines regions where the surface modes can exist.  The existence of an imaginary part indicates that the solution in Eq.(\ref{mat}) is a psuedosurface mode, and not purely localized to the surface. Instead, energy "leaks" into the bulk. The wave is comprised of a localised component, that travels along the surface and decays into the material according
to the real part of  $\beta $, and a component that travels into the material with wave number equal to the imaginary part of \textit{${\beta}$}. 

Values for the imaginary parts of  $\beta $ are
plotted as a function of frequency in Fig.\ref{imagbeta} for the case of no applied fields. Imaginary  $\beta $  depends linearly  on the   magnetoelectric susceptibility and becomes large near the electric resonance frequency  $\acute{\omega
}_{\mathit{ey}}$. One can also see from Fig.\ref{imagbeta}, that the coupling directly influences the magnitude of "leakage". If the coupling is large, then the ME susceptibility will also be large and thereby increase the imaginary part of $\beta $. 

In the case where $k >> \frac{\omega}{c}$, the attenuation constants for the material and vacuum regions can be approximated by  $\epsilon_z\beta^2\approx \epsilon_{y}k_y^2$ and $\beta_o\approx k_y$.  The dispersion relation Eq.(\ref{bulk}) then reduces to:
\begin{equation}
\label{magnetostatic}
	\epsilon_z\epsilon_y = 1.
\end{equation}
The surface modes require the permitivity $\epsilon_y$ to be negative, and the longitudinal phonon frequency polarized along $z$ direction, $\omega_{oz}$, is lower than $\omega'_{ey}$. This means that the value of $\epsilon_z$ is positive in  regions where  surface modes exist, and the requirement in Eq.(\ref{magnetostatic}) is never satisfied.  Therefore, surface modes do not exist in the limit $k >> \frac{\omega}{c}$.


\begin{figure}[ht]
\begin{center}
\subfigure[\label{esdt}E vs angle.]{
\includegraphics[width=6.4cm]{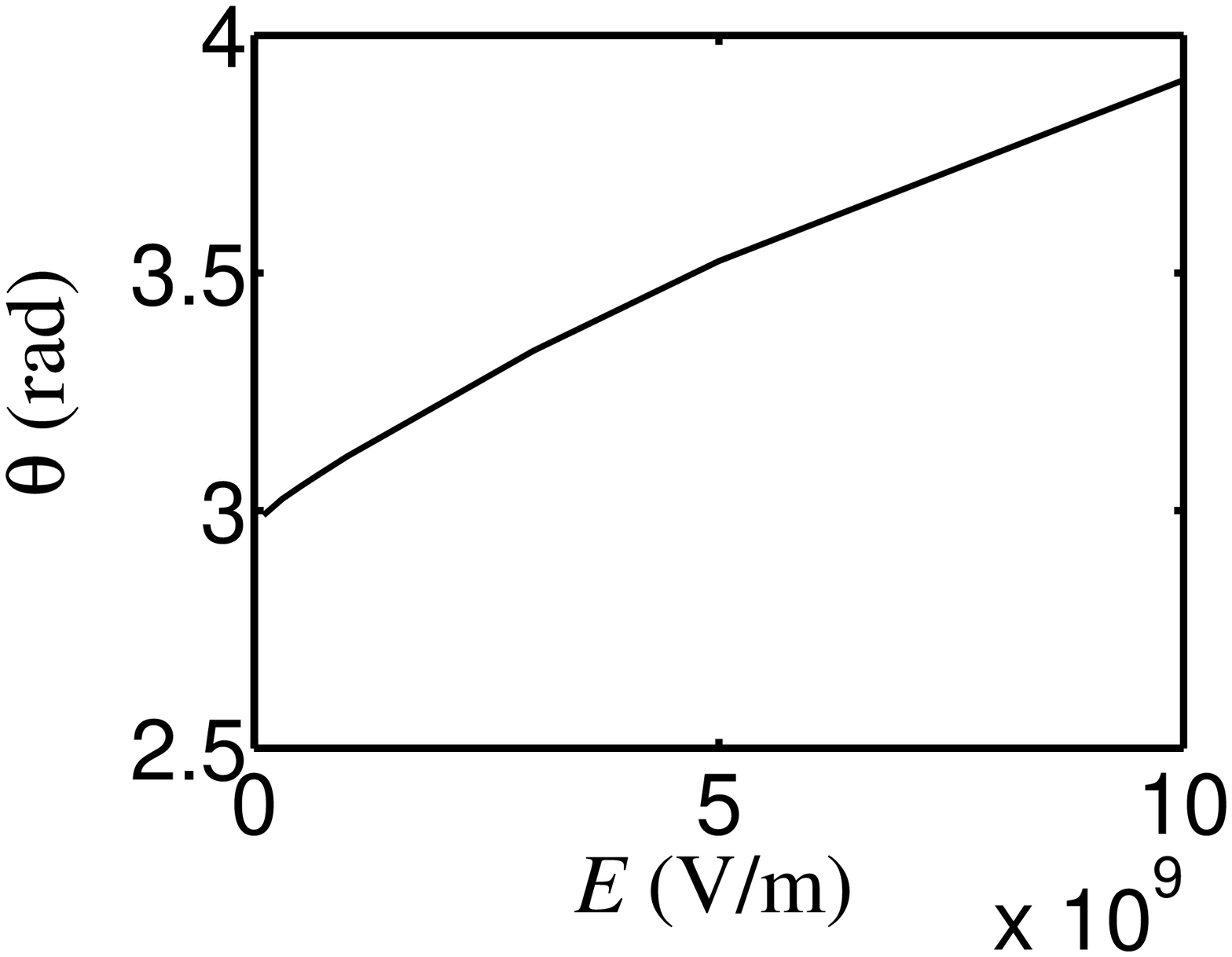}}
\subfigure[\label{reldispE}Influence of E into Dispersion relation.]{
\includegraphics[width=7cm]{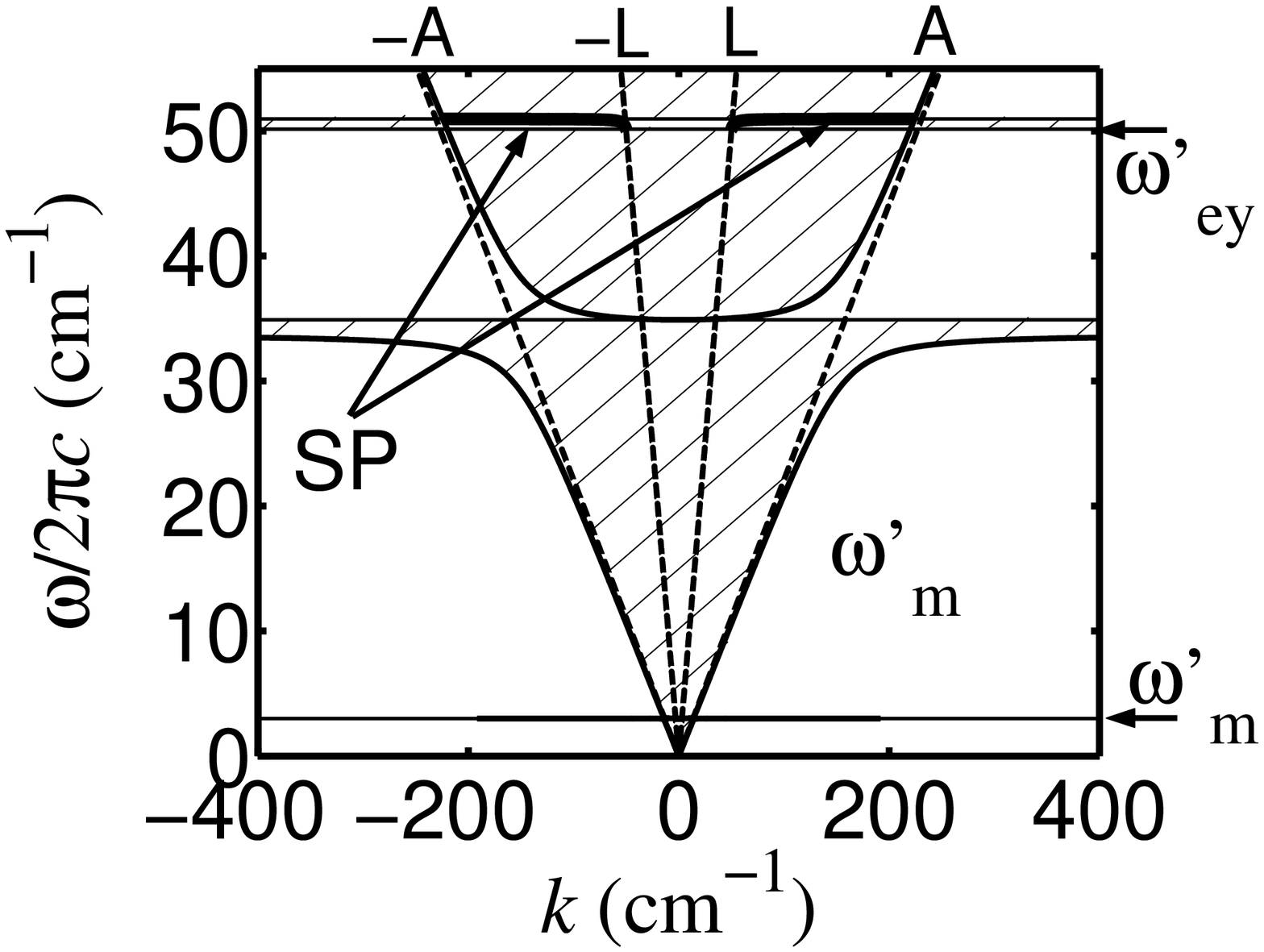}}
\subfigure[\label{hosdt}H$_o$ vs angle.]{
\includegraphics[width=6.4cm,height=4.9cm]{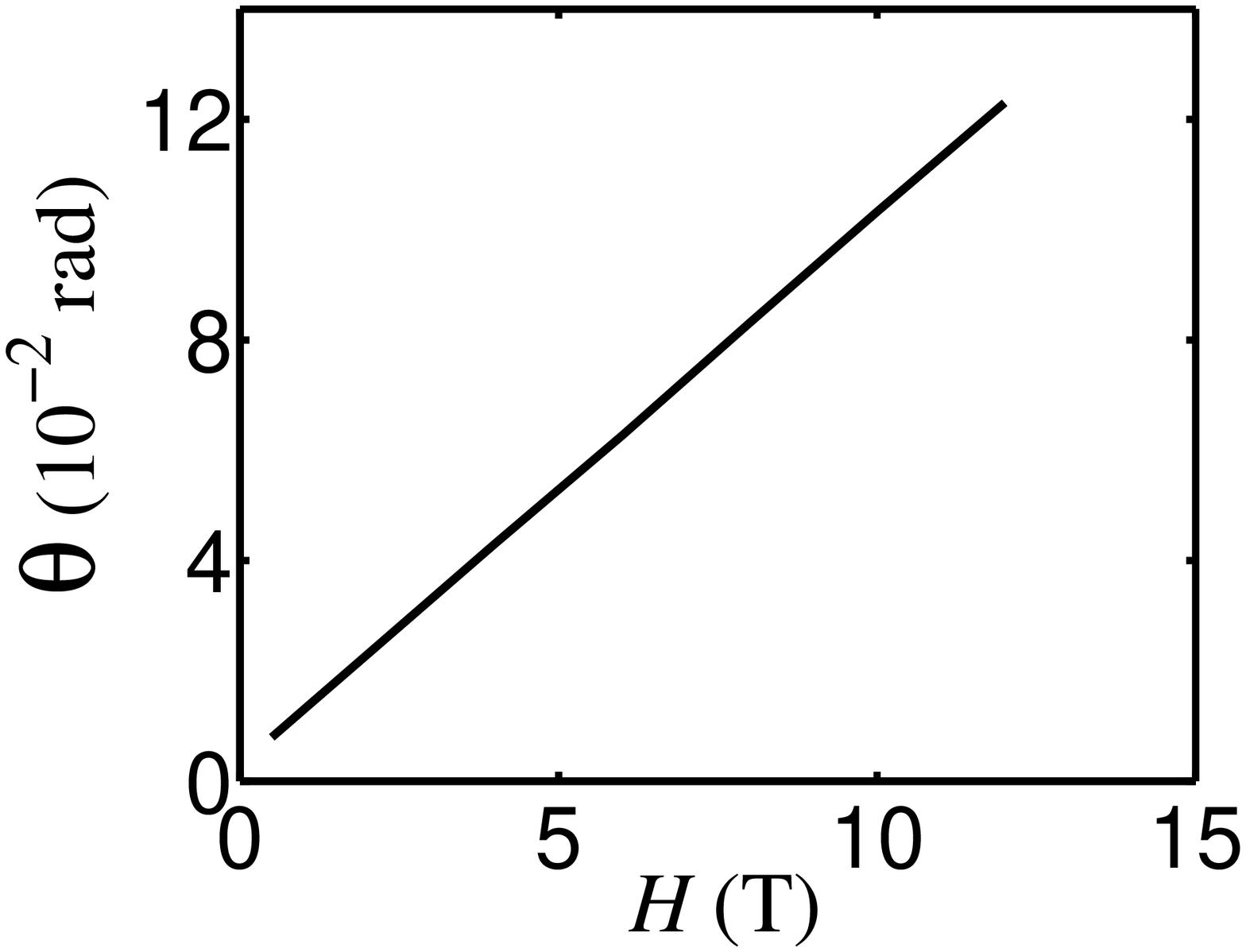}}
\subfigure[\label{reldispHo}Influence of H into Dispersion relation.]{
\includegraphics[width=7.4cm,height=5.3cm]{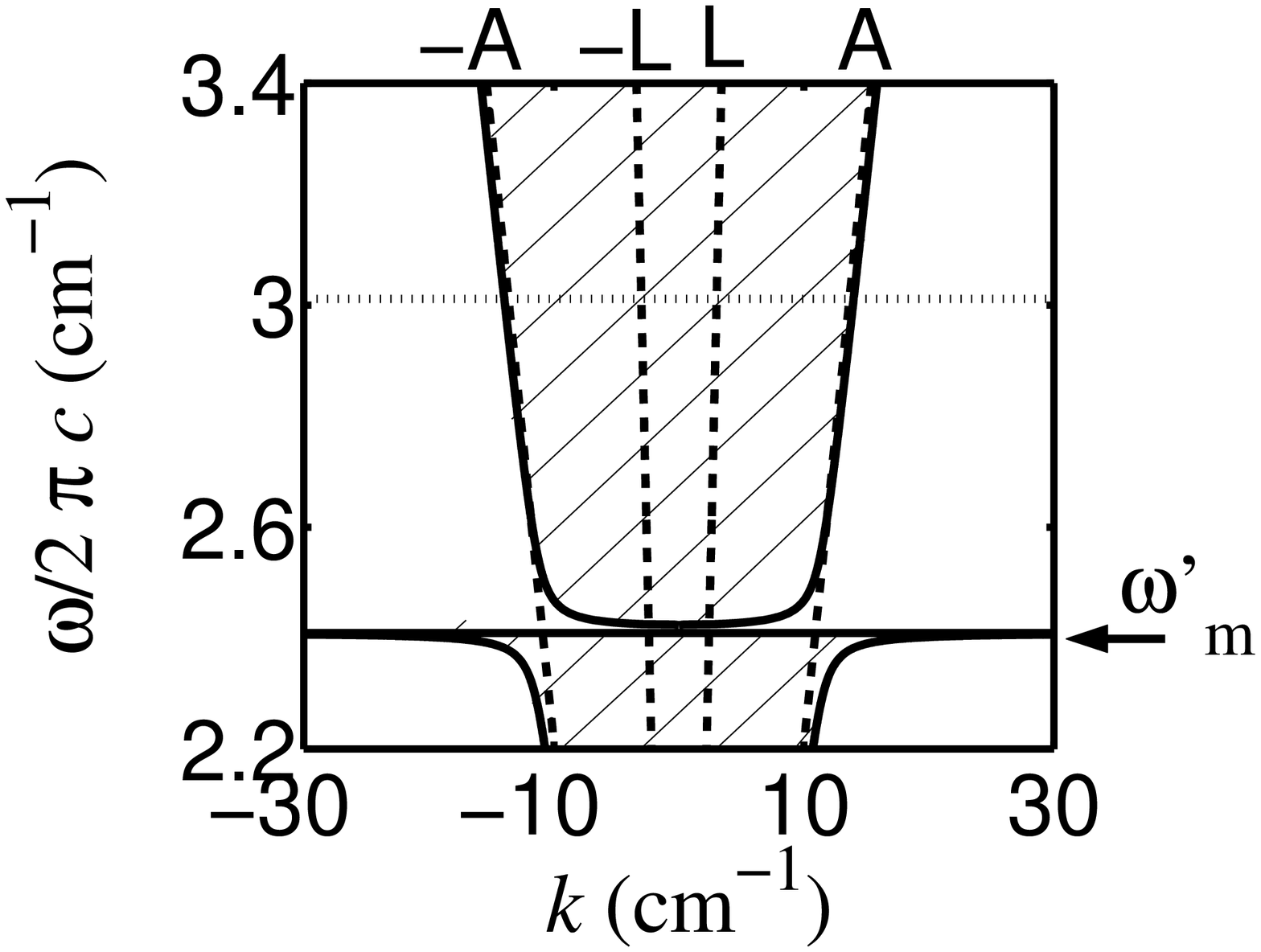}}
\caption{Influence external field in the canting angle and dispersion relation  In (a)the external electric field E increase the canting angle with small value. In (b) dispersion relation is shown with an external electric field along the spontaneous polarisation ${E} = 5\times10^8$ V/m. In (c)the external electric field H$_o$ increase the canting angle. In (d), magnetic resonance frequency shift down when the external magnetic field, $H_{o} = 10$T is applied. The dotted line is the magnetic frequency in the absence of magnetic field.}
\end{center}
\end{figure}

We now study the influence of external fields on the band structures. Results are shown in Figs.\ref{reldispE} and \ref{reldispHo}. Results for an electric field value of 5${\times}$ 10\textsuperscript{8} V/m along the direction of spontaneous polarisation (but with zero magnetic field) are shown in Fig.\ref{reldispE}. The effect of the electric field is to shift the window where  surface modes exist to higher frequencies. The upward frequency shift is due to the increase of spontaneous polarisation, which directly increases the phonon frequencies $\omega'_{ey}$ and $\omega_{Ly}$.  However, the change in $\omega_{Ly}$ is smaller than the change in $\omega'_{ey}$ (which is due to the third term in Eq.(\ref{WL})) and so the surface mode window is narrowed.  The electric field increases the canting angle slightly, as shown in Fig.\ref{esdt}, and the effect on magnetic resonance is negligible.

Results for a  magnetic field of 10 T (with zero electric field) are shown in Fig.\ref{reldispHo}. The magnetic field increases the canting angle (see Fig.\ref{hosdt}) and shifts  $\acute{\omega}_{m}$ to a lower frequency (as shown in Fig.\ref{reldispHo}) but the effects on the bulk bands are negligible. This shift can be understood if we consider the case where the ME coupling is neglected.  Then the frequency $\omega_{me}\longrightarrow0$, and also $\Omega_{me} \longrightarrow0$. In this case, the magnetic resonance frequency will take the form $\omega^2_m=\tilde{\omega}^2_{afm}+\Omega^2_0\longrightarrow\omega_a(\omega_a-2\omega_ex)\cos^2{\theta}$.  It can be seen from this expression that a magnetic field reduces the magnetic resonance frequency.  We note that this effect is  mentioned in Ref.\cite{almeida}.

\begin{figure}[ht]
\begin{center}
\subfigure[\label{reldispf}Dispersion relation with close frequency of magnetic and electric]{
\includegraphics[width=7cm]{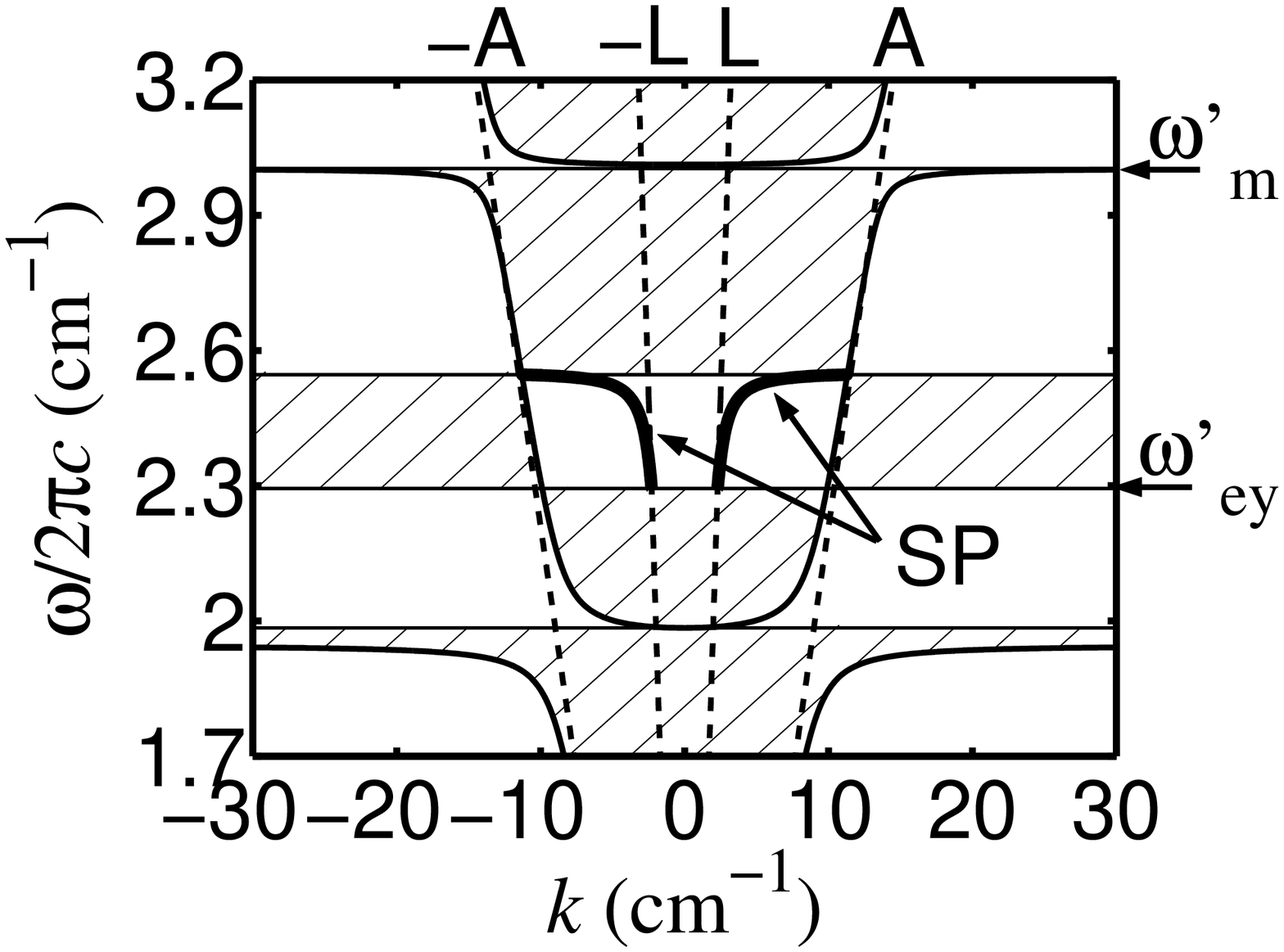}}
\subfigure[\label{reldispfho}The magnetic resonance inside the window]{
\includegraphics[width=7cm]{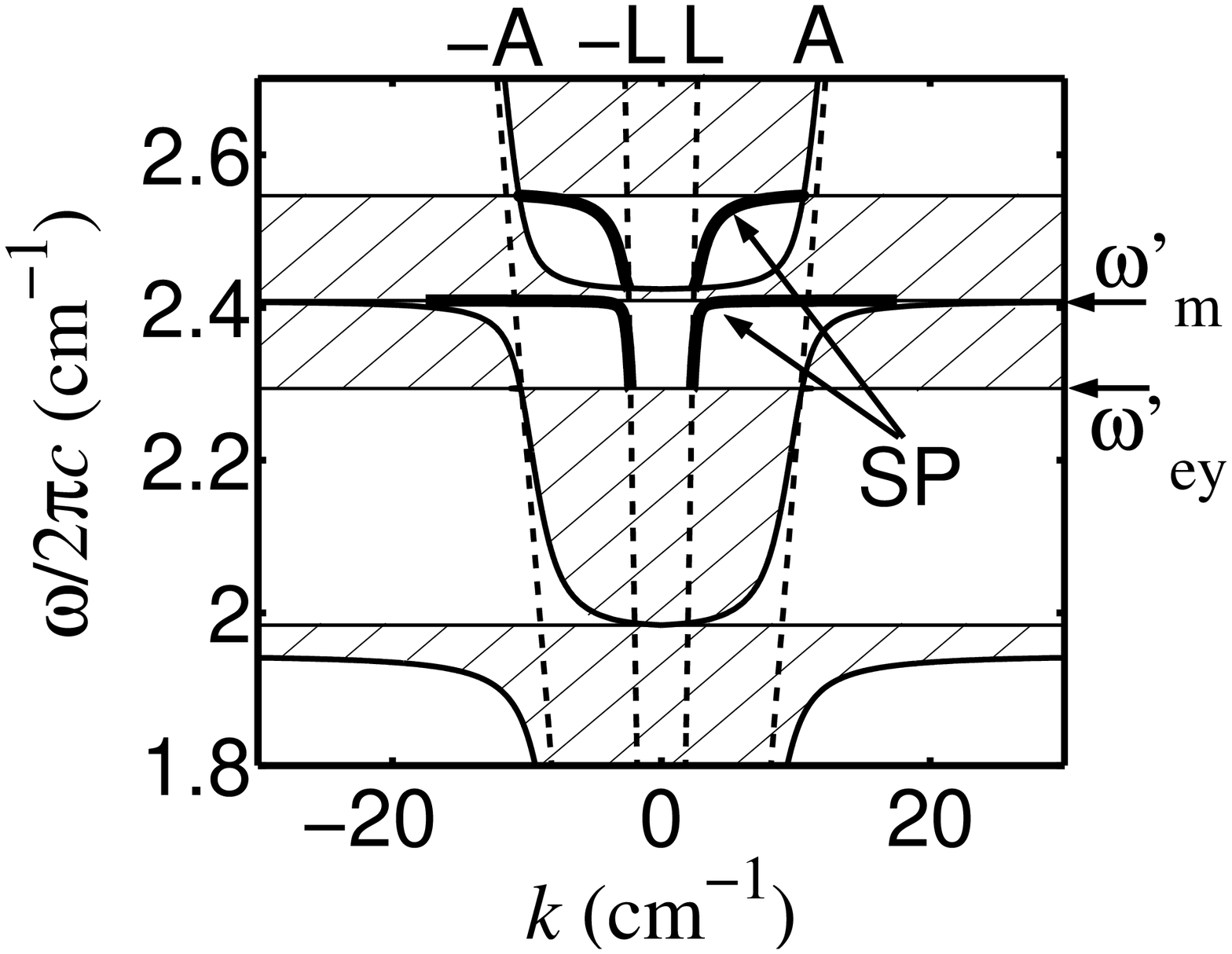}}
\caption{Dispersion relation for a material with  electric and magnetic resonances near one another in frequency.  In (a) the dispersion relation without external magnetic field is shown. In (b), the magnetic resonance frequency is shifted into a surface mode window with the application of a large magnetic field, $H_{o} = 12$T.}
\end{center}
\end{figure}

The surface modes can be modified if the magnetic resonance frequency can be shifted into a surface wave ``window'',  as shown in Fig.\ref{reldispfho} and \ref{T250}. In BaMnF$_{4}$ where the electric and magnetic resonance frequency are well separated around 37cm$^{-1}$, it is very difficult to arrange the magnetic resonance frequency inside this window. This may be possible for a suitably prepared material (or artificially constructed material) whose  frequency separation between electric and magnetic resonance is smaller.

\section{Parameter effects on surface mode properties}
Lastly, we identify the key parameters affecting surface mode frequencies. In the first case, changing the phonon mass to $f = 9\times{10^{21}}$statA$^{2}$s$^{2}$/gcm$^{3}$ moves the magnetic resonance frequency to 1 cm$^{-1}$ above the the electric resonance frequency $\omega_{ey}$. The result on surface and bulk polariton bands is shown in Fig.\ref{reldispf}. The dielectric constant background has also been reduced to $\epsilon^{\infty}_y=2.6$, in order to widen the surface mode window. Application of an external magnetic field lowers the magnetic resonance frequency.  Application of a large external magnetic of 12 T places the magnetic resonance inside the window as illustrated in Fig.\ref{reldispfho}.

Inside the window, the magnetic resonance splits the surface mode into low and high frequency branches for each direction of propagation. The properties of the upper part are similar to that discussed in the previous section. However, the lower branch terminates at the magnetic resonance frequency as illustrated in Fig.\ref{reldispfho}.  In this case, the requirement that the dielectric constant $\epsilon_y$ should be negative for surface modes prevents both the upper and lower branches to exist in the region where $k>>\frac{\omega}{c}$.  

\begin{figure}[ht]
\begin{center}
\subfigure[\label{T150}Dispersion relation with by modifying exchange constant at T=150 K]{
\includegraphics[width=7cm]{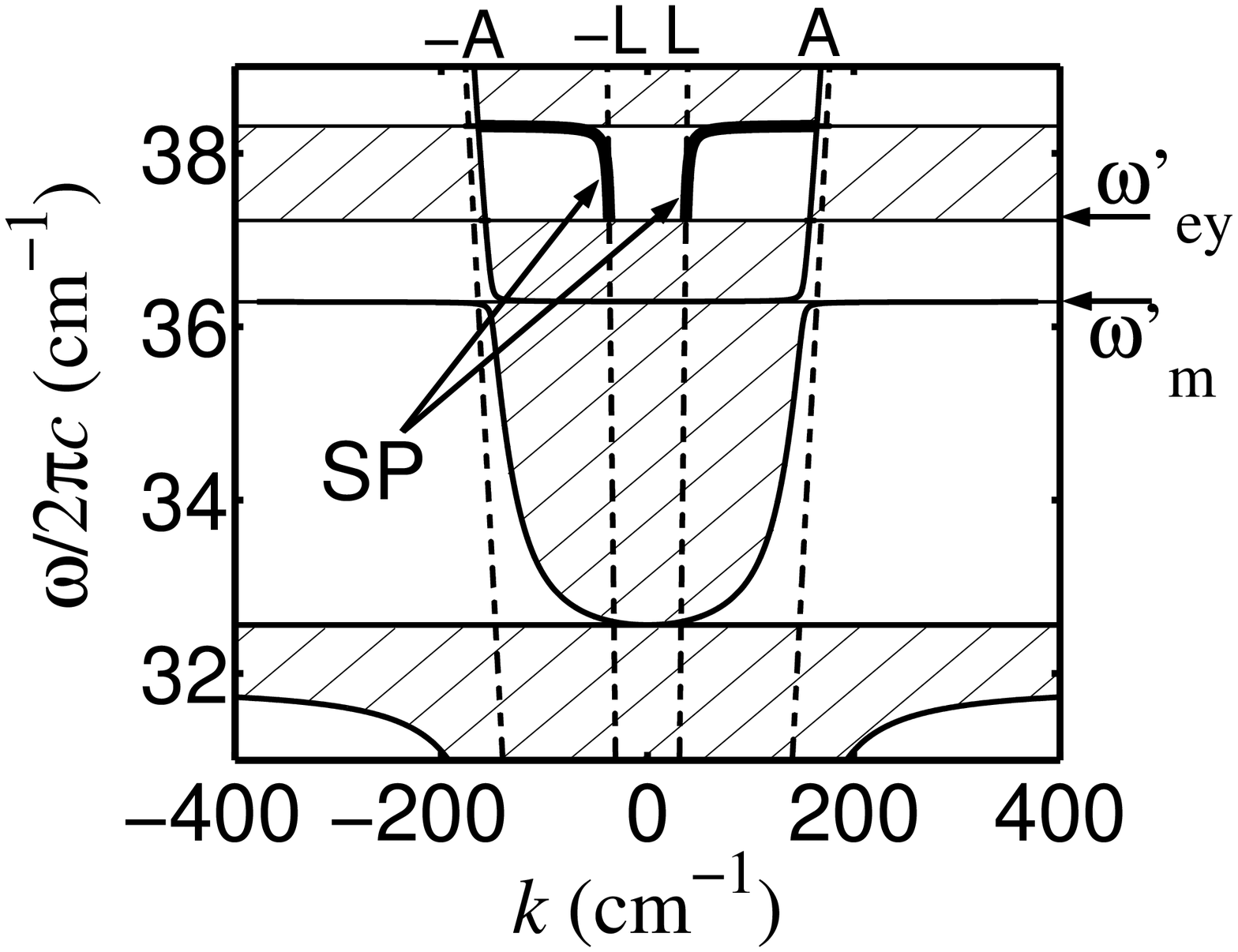}}
\subfigure[\label{T250}Dispersion relation at T=250 K]{
\includegraphics[width=7cm]{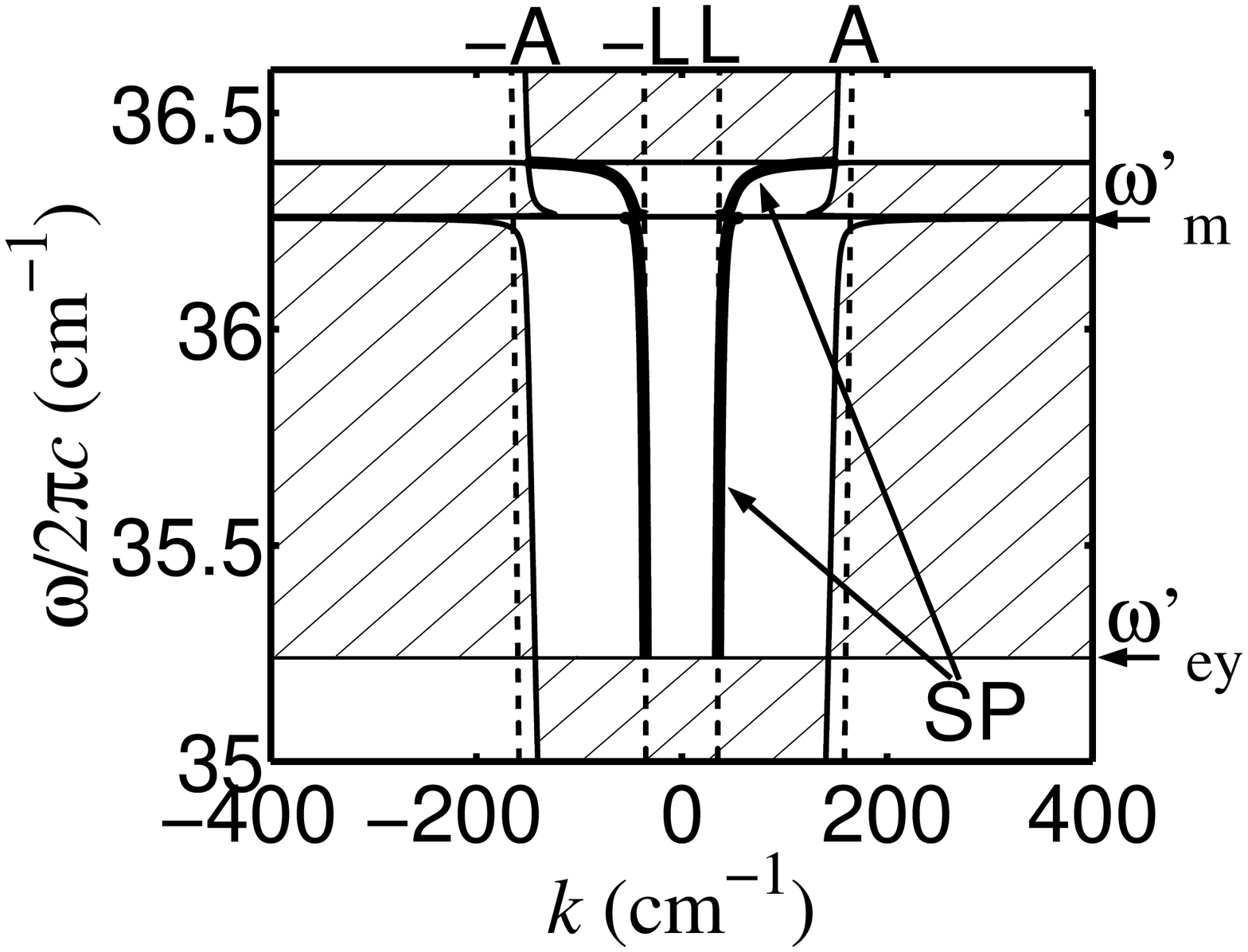}}
\caption{Temperature dependence with modified exchange constant. (a) Dispersion relation at T=150 K. In (b), the magnetic resonance frequency shifts down at 250 K}
\end{center}
\end{figure}

 In a second example, we consider if the exchange constant is -4000 and the anisotropy constant is  4. The ME coupling is also changed to 40 m$^{2}$/C, which keeps the canting angle small.  The dispersion relation at temperature 150 K is presented in Fig.\ref{T150}.  As temperature is increased to 250 K, the polarisation will decrease while the magnetisation does not change significantly. Hence, the  electric resonance goes below the magnetic resonance frequency.  Results are shown in Fig.\ref{T250}. As in the first example, the magnetic resonance frequency then exists inside the surface mode window and a similar splitting of the surface mode branches occurs.


\section{Conclusions}
\label{Conclusions}
We have shown how linear magnetoelectric coupling influences surface and bulk TM polaritons modes in a canted multiferroic. In the first place, a narrow restrahl region forms in the bulk mode band at a frequency near the longitudinal phonon frequency (along the spontaneous polarization).  Reciprocal surface excitations can exist in this region. In general the surface exciations in this polarization are actually psuedosurface waves that are only partially localized to the surface. The imaginary part of the decay constant is proportional to the magneto-electric coupling.

We have explored also possibilities of modifying the polariton band structure through changes in key material parameters. Such effects may possibly be realized in appropriate compounds, or in artifically layered heterostructures. If the magnetoelectric material has a magnetic resonance frequency above the electric resonance frequency, and the difference between those frequencies is not great, then it is possible to create surface excitations that are sensitive to temperature, electric fields, and magnetic fields.

\begin{acknowledgments}
We wish to acknowledge the support of Ausaid, the Australian Research Council and DEST.
\end{acknowledgments}


\begin{thebibliography}{10}%
\makeatletter
\providecommand \@ifxundefined [1]{%
 \ifx #1\undefined \expandafter \@firstoftwo
 \else \expandafter \@secondoftwo
\fi
}%
\providecommand \@ifnum [1]{%
 \ifnum #1\expandafter \@firstoftwo
 \else \expandafter \@secondoftwo
\fi
}%
\providecommand \enquote [1]{``#1''}%
\providecommand \bibnamefont  [1]{#1}%
\providecommand \bibfnamefont [1]{#1}%
\providecommand \citenamefont [1]{#1}%
\providecommand\href[0]{\@sanitize\@href}%
\providecommand\@href[1]{\endgroup\@@startlink{#1}\endgroup\@@href}%
\providecommand\@@href[1]{#1\@@endlink}%
\providecommand \@sanitize [0]{\begingroup\catcode`\&12\catcode`\#12\relax}%
\@ifxundefined \pdfoutput {\@firstoftwo}{%
 \@ifnum{\z@=\pdfoutput}{\@firstoftwo}{\@secondoftwo}%
}{%
 \providecommand\@@startlink[1]{\leavevmode}%
 \providecommand\@@endlink[0]{}%
}{%
 \providecommand\@@startlink[1]{%
  \leavevmode
  \pdfstartlink
   attr{/Border[0 0 1 ]/H/I/C[0 1 1]}%
   user{/Subtype/Link/A<</Type/Action/S/URI/URI(#1)>>}%
  \relax
 }%
 \providecommand\@@endlink[0]{\pdfendlink}%
}%
\providecommand \url  [0]{\begingroup\@sanitize \@url }%
\providecommand \@url [1]{\endgroup\@href {#1}{\urlprefix}}%
\providecommand \urlprefix [0]{URL }%
\providecommand \Eprint[0]{\href }%
\@ifxundefined \urlstyle {%
  \providecommand \doi [1]{doi:\discretionary{}{}{}#1}%
}{%
  \providecommand \doi [0]{doi:\discretionary{}{}{}\begingroup
  \urlstyle{rm}\Url }%
}%
\providecommand \doibase [0]{http://dx.doi.org/}%
\providecommand \Doi[1]{\href{\doibase#1}}%
\providecommand \bibAnnote [3]{%
  \BibitemShut{#1}%
  \begin{quotation}\noindent
    \textsc{Key:}\ #2\\\textsc{Annotation:}\ #3%
  \end{quotation}%
}%
\providecommand \bibAnnoteFile [2]{%
  \IfFileExists{#2}{\bibAnnote {#1} {#2} {\input{#2}}}{}%
}%
\providecommand \typeout [0]{\immediate \write \m@ne }%
\providecommand \selectlanguage [0]{\@gobble}%
\providecommand \bibinfo [0]{\@secondoftwo}%
\providecommand \bibfield [0]{\@secondoftwo}%
\providecommand \translation [1]{[#1]}%
\providecommand \BibitemOpen[0]{}%
\providecommand \bibitemStop [0]{}%
\providecommand \bibitemNoStop [0]{.\EOS\space}%
\providecommand \EOS [0]{\spacefactor3000\relax}%
\providecommand \BibitemShut [1]{\csname bibitem#1\endcsname}%
\bibitem{camley82}%
  \BibitemOpen
  \bibfield{author}{%
  \bibinfo {author} {\bibfnamefont{R.~E.}\ \bibnamefont{Camley}}\ and\ \bibinfo
  {author} {\bibfnamefont{D.~L.}\ \bibnamefont{Mills}},\ }%
  \bibfield{journal}{%
  \bibinfo {journal} {Phys. Rev.}\ }%
  \textbf{\bibinfo {volume} {26}},\ \bibinfo {pages} {1280} (\bibinfo {year}
  {1982})%
  \bibAnnoteFile{NoStop}{camley82}%
\bibitem{harstein73}%
  \BibitemOpen
  \bibfield{author}{%
  \bibinfo {author} {\bibfnamefont{A.}~\bibnamefont{Harstein}}, \bibinfo
  {author} {\bibfnamefont{E.}~\bibnamefont{Burstein}}, \bibinfo {author}
  {\bibfnamefont{A.~A.}\ \bibnamefont{Maradudin}}, \bibinfo {author}
  {\bibfnamefont{R.}~\bibnamefont{Brewer}},\ and\ \bibinfo {author}
  {\bibfnamefont{J.~F.}\ \bibnamefont{Wallis}},\ }%
  \bibfield{journal}{%
  \bibinfo {journal} {J. Phys. C: Solid State Physics}\ }%
  \textbf{\bibinfo {volume} {6}},\ \bibinfo {pages} {1266} (\bibinfo {year}
  {1973})%
  \bibAnnoteFile{NoStop}{harstein73}%
\bibitem{boardman}%
  \BibitemOpen
  \bibfield{author}{%
  \bibinfo {author} {\bibfnamefont{E.~F.}\ \bibnamefont{Sarmento}}\ and\
  \bibinfo {author} {\bibfnamefont{D.~R.}\ \bibnamefont{Tilley}},\ }%
  \emph{\bibinfo {title} {Electromagnetic surface modes}}\ (\bibinfo
  {publisher} {John Wiley and sons},\ \bibinfo {year} {1982})\ p.\ \bibinfo
  {pages} {633}%
  \bibAnnoteFile{NoStop}{boardman}%
\bibitem{barnas86a}%
  \BibitemOpen
  \bibfield{author}{%
  \bibinfo {author} {\bibfnamefont{J.}~\bibnamefont{Barnas}},\ }%
  \bibfield{journal}{%
  \bibinfo {journal} {J. Magn. Magn. Mat.}\ }%
  \textbf{\bibinfo {volume} {62}},\ \bibinfo {pages} {381} (\bibinfo {year}
  {1986})%
  \bibAnnoteFile{NoStop}{barnas86a}%
\bibitem{nylander}%
  \BibitemOpen
  \bibfield{author}{%
  \bibinfo {author} {\bibfnamefont{C.}~\bibnamefont{Nylander}}, \bibinfo
  {author} {\bibfnamefont{B.}~\bibnamefont{Liedberg}},\ and\ \bibinfo {author}
  {\bibfnamefont{T.}~\bibnamefont{Lind}},\ }%
  \bibfield{journal}{%
  \bibinfo {journal} {Sens. and Actuators}\ }%
  \textbf{\bibinfo {volume} {3}},\ \bibinfo {pages} {79} (\bibinfo {year}
  {1982})%
  \bibAnnoteFile{NoStop}{nylander}%
\bibitem{liedberg}%
  \BibitemOpen
  \bibfield{author}{%
  \bibinfo {author} {\bibfnamefont{B.}~\bibnamefont{Liedberg}}, \bibinfo
  {author} {\bibfnamefont{C.}~\bibnamefont{Nylander}},\ and\ \bibinfo {author}
  {\bibfnamefont{I.}~\bibnamefont{Lundstorm}},\ }%
  \bibfield{journal}{%
  \bibinfo {journal} {Sens. and Actuators}\ }%
  \textbf{\bibinfo {volume} {4}},\ \bibinfo {pages} {299} (\bibinfo {year}
  {1983})%
  \bibAnnoteFile{NoStop}{liedberg}%
\bibitem{keilmann98}%
  \BibitemOpen
  \bibfield{author}{%
  \bibinfo {author} {\bibfnamefont{F.}~\bibnamefont{Keilmann}},\ }%
  \bibfield{journal}{%
  \bibinfo {journal} {J. Micros.}\ }%
  \textbf{\bibinfo {volume} {194}},\ \bibinfo {pages} {567} (\bibinfo {year}
  {1999})%
  \bibAnnoteFile{NoStop}{keilmann98}%
\bibitem{kars78}%
  \BibitemOpen
  \bibfield{author}{%
  \bibinfo {author} {\bibfnamefont{A.~D.}\ \bibnamefont{Karsono}}\ and\
  \bibinfo {author} {\bibfnamefont{D.~R.}\ \bibnamefont{Tilley}},\ }%
  \bibfield{journal}{%
  \bibinfo {journal} {J. Phys. C.}\ }%
  \textbf{\bibinfo {volume} {11}},\ \bibinfo {pages} {3487} (\bibinfo {year}
  {1978})%
  \bibAnnoteFile{NoStop}{kars78}%
\bibitem{abraha96}%
  \BibitemOpen
  \bibfield{author}{%
  \bibinfo {author} {\bibfnamefont{K.}~\bibnamefont{Abraha}}\ and\ \bibinfo
  {author} {\bibfnamefont{D.~R.}\ \bibnamefont{Tilley}},\ }%
  \bibfield{journal}{%
  \bibinfo {journal} {Surf. Sci. Rep.}\ }%
  \textbf{\bibinfo {volume} {24}},\ \bibinfo {pages} {129} (\bibinfo {year}
  {1996})%
  \bibAnnoteFile{NoStop}{abraha96}%
\bibitem{camley98}%
  \BibitemOpen
  \bibfield{author}{%
  \bibinfo {author} {\bibfnamefont{R.~E.}\ \bibnamefont{Camley}}, \bibinfo
  {author} {\bibfnamefont{M.~R.~F.}\ \bibnamefont{Jensen}}, \bibinfo {author}
  {\bibfnamefont{S.~A.}\ \bibnamefont{Feiven}},\ and\ \bibinfo {author}
  {\bibfnamefont{T.~J.}\ \bibnamefont{Parker}},\ }%
  \bibfield{journal}{%
  \bibinfo {journal} {J. Appl. Phys.}\ }%
  \textbf{\bibinfo {volume} {83}},\ \bibinfo {pages} {6280} (\bibinfo {year}
  {1998})%
  \bibAnnoteFile{NoStop}{camley98}%
\bibitem{barnas86b}%
  \BibitemOpen
  \bibfield{author}{%
  \bibinfo {author} {\bibfnamefont{J.}~\bibnamefont{Barnas}},\ }%
  \bibfield{journal}{%
  \bibinfo {journal} {J. Phys. C: Solid State Physics}\ }%
  \textbf{\bibinfo {volume} {19}},\ \bibinfo {pages} {419} (\bibinfo {year}
  {1986})%
  \bibAnnoteFile{NoStop}{barnas86b}%
\bibitem{tarasenko00}%
  \BibitemOpen
  \bibfield{author}{%
  \bibinfo {author} {\bibfnamefont{S.~V.}\ \bibnamefont{Tarasenko}}\ and\
  \bibinfo {author} {\bibfnamefont{V.~G.}\ \bibnamefont{Shavrov}},\ }%
  \bibfield{journal}{%
  \bibinfo {journal} {Ferroelectrics}\ }%
  \textbf{\bibinfo {volume} {279}},\ \bibinfo {pages} {3} (\bibinfo {year}
  {2002})%
  \bibAnnoteFile{NoStop}{tarasenko00}%
\bibitem{buchel86}%
  \BibitemOpen
  \bibfield{author}{%
  \bibinfo {author} {\bibfnamefont{V.~D.}\ \bibnamefont{Buchel'nikov}}\ and\
  \bibinfo {author} {\bibfnamefont{V.~G.}\ \bibnamefont{Shavrov}},\ }%
  \bibfield{journal}{%
  \bibinfo {journal} {JETP}\ }%
  \textbf{\bibinfo {volume} {82}},\ \bibinfo {pages} {380} (\bibinfo {year}
  {1996})%
  \bibAnnoteFile{NoStop}{buchel86}%
\bibitem{tilley82}%
  \BibitemOpen
  \bibfield{author}{%
  \bibinfo {author} {\bibfnamefont{D.~R.}\ \bibnamefont{Tilley}}\ and\ \bibinfo
  {author} {\bibfnamefont{J.~F.}\ \bibnamefont{Scott}},\ }%
  \bibfield{journal}{%
  \bibinfo {journal} {Phys. Rev. B}\ }%
  \textbf{\bibinfo {volume} {25}},\ \bibinfo {pages} {3251} (\bibinfo {year}
  {1982})%
  \bibAnnoteFile{NoStop}{tilley82}%
\bibitem{ederer08}%
  \BibitemOpen
  \bibfield{author}{%
  \bibinfo {author} {\bibfnamefont{C.}~\bibnamefont{Ederer}}\ and\ \bibinfo
  {author} {\bibfnamefont{C.~J.}\ \bibnamefont{Fennie}},\ }%
  \bibfield{journal}{%
  \bibinfo {journal} {J. Phys.: Condens Matter}\ }%
  \textbf{\bibinfo {volume} {20}},\ \bibinfo {pages} {434219} (\bibinfo {year}
  {2008})%
  \bibAnnoteFile{NoStop}{ederer08}%
\bibitem{scott79}%
  \BibitemOpen
  \bibfield{author}{%
  \bibinfo {author} {\bibfnamefont{J.}~\bibnamefont{Scott}},\ }%
  \bibfield{journal}{%
  \bibinfo {journal} {Rep. Prog. Phys.}\ }%
  \textbf{\bibinfo {volume} {12}},\ \bibinfo {pages} {1055} (\bibinfo {year}
  {1979})%
  \bibAnnoteFile{NoStop}{scott79}%
\bibitem{venturini}%
  \BibitemOpen
  \bibfield{author}{%
  \bibinfo {author} {\bibfnamefont{E.~L.}\ \bibnamefont{Venturini}}\ and\
  \bibinfo {author} {\bibfnamefont{F.~R.}\ \bibnamefont{Morgenthaler}},\ }%
  \bibfield{journal}{%
  \bibinfo {journal} {AIP Conf. Proc.}\ }%
  \textbf{\bibinfo {volume} {24}},\ \bibinfo {pages} {168} (\bibinfo {year}
  {1975})%
  \bibAnnoteFile{NoStop}{venturini}%
\bibitem{holmes69}%
  \BibitemOpen
  \bibfield{author}{%
  \bibinfo {author} {\bibfnamefont{L.}~\bibnamefont{Holmes}}, \bibinfo {author}
  {\bibfnamefont{M.}~\bibnamefont{Eibschutz}},\ and\ \bibinfo {author}
  {\bibfnamefont{H.~J.}\ \bibnamefont{Guggenheim}},\ }%
  \bibfield{journal}{%
  \bibinfo {journal} {Solid State Commun}\ }%
  \textbf{\bibinfo {volume} {7}},\ \bibinfo {pages} {973} (\bibinfo {year}
  {1969})%
  \bibAnnoteFile{NoStop}{holmes69}%
\bibitem{samara76}%
  \BibitemOpen
  \bibfield{author}{%
  \bibinfo {author} {\bibfnamefont{G.~A.}\ \bibnamefont{Samara}}\ and\ \bibinfo
  {author} {\bibfnamefont{P.~M.}\ \bibnamefont{Richards}},\ }%
  \bibfield{journal}{%
  \bibinfo {journal} {Phys. Rev. B}\ }%
  \textbf{\bibinfo {volume} {14}},\ \bibinfo {pages} {5073} (\bibinfo {year}
  {1976})%
  \bibAnnoteFile{NoStop}{samara76}%
\bibitem{keve71}%
  \BibitemOpen
  \bibfield{author}{%
  \bibinfo {author} {\bibfnamefont{S.~C.}\ \bibnamefont{Abrahams}}\ and\
  \bibinfo {author} {\bibfnamefont{E.~T.}\ \bibnamefont{Keve}},\ }%
  \bibfield{journal}{%
  \bibinfo {journal} {Ferroelectrics}\ }%
  \textbf{\bibinfo {volume} {2}},\ \bibinfo {pages} {129} (\bibinfo {year}
  {1971})%
  \bibAnnoteFile{NoStop}{keve71}%
\bibitem{lim}%
  \BibitemOpen
  \bibfield{author}{%
  \bibinfo {author} {\bibfnamefont{T.~C.}\ \bibnamefont{Lim}}\ and\ \bibinfo
  {author} {\bibfnamefont{G.~W.}\ \bibnamefont{Farnell}},\ }%
  \bibfield{journal}{%
  \bibinfo {journal} {J. Appl. Phys.}\ }%
  \textbf{\bibinfo {volume} {39}},\ \bibinfo {pages} {4319} (\bibinfo {year}
  {1968})%
  \bibAnnoteFile{NoStop}{lim}%
\bibitem{almeida}%
  \BibitemOpen
  \bibfield{author}{%
  \bibinfo {author} {\bibfnamefont{N.~S.}\ \bibnamefont{Almeida}}\ and\
  \bibinfo {author} {\bibfnamefont{D.~L.}\ \bibnamefont{Mills}},\ }%
  \bibfield{journal}{%
  \bibinfo {journal} {Phys. Rev. B.}\ }%
  \textbf{\bibinfo {volume} {37}},\ \bibinfo {pages} {3400} (\bibinfo {year}
  {1988})%
  \bibAnnoteFile{NoStop}{almeida}%
\end{thebibliography}
\end{document}